\documentclass[twocolumn]{aastex63}

\usepackage{color}
\usepackage{graphicx,graphics}
\usepackage{amssymb,amsmath}
\usepackage{natbib}
\usepackage {threeparttable}
\usepackage {bm}

\newcommand{\Mstar}{M_\star}

\shorttitle{Environmental dependence of galactic properties traced by Ly$\alpha$ forest absorption}
\shortauthors{R. Momose et al.}

\begin{document}

\title{Environmental Dependence of Galactic Properties Traced by Ly$\alpha$ Forest Absorption: \\
Diversity among Galaxy Populations}


\correspondingauthor{Rieko Momose}
\email{momose@astron.s.u-tokyo.ac.jp}

\author[0000-0002-8857-2905]{Rieko Momose}
\affiliation{Department of Astronomy, School of Science, The University of Tokyo, 
7-3-1 Hongo, Bunkyo-ku, Tokyo 113-0033, Japan}

\author[0000-0002-2597-2231]{Kazuhiro Shimasaku}
\affiliation{Department of Astronomy, School of Science, The University of Tokyo, 
7-3-1 Hongo, Bunkyo-ku, Tokyo 113-0033, Japan}
\affiliation{Research Center for the Early Universe, The University of Tokyo, 7-3-1 Hongo, Bunkyo-ku, Tokyo 113-0033, Japan}

\author[0000-0003-3954-4219]{Nobunari Kashikawa}
\affiliation{Department of Astronomy, School of Science, The University of Tokyo, 
7-3-1 Hongo, Bunkyo-ku, Tokyo 113-0033, Japan}
\affiliation{Research Center for the Early Universe, The University of Tokyo, 7-3-1 Hongo, Bunkyo-ku, Tokyo 113-0033, Japan}

\author[0000-0001-7457-8487]{Kentaro Nagamine}
\affiliation{Theoretical Astrophysics, Department of Earth and Space Science, Osaka University, 1-1 Machikaneyama, \\
Toyonaka, Osaka 560-0043, Japan}
\affiliation{Department of Physics and Astronomy, University of Nevada, Las Vegas, NV 89154-4002, USA}
\affiliation{Kavli-IPMU (WPI), The University of Tokyo, 5-1-5 Kashiwanoha, Kashiwa, Chiba 277-8583, Japan}

\author{Ikkoh Shimizu}
\affiliation{Shikoku Gakuin University, 3-2-1 Bunkyocho, Zentsuji, Kagawa 765-0013, Japan}
\affiliation{National Astronomical Observatory of Japan, 2-21-1 Osawa, Mitaka, Tokyo 181-8588, Japan}

\author{Kimihiko Nakajima}
\affiliation{National Astronomical Observatory of Japan, 2-21-1 Osawa, Mitaka, Tokyo 181-8588, Japan}
\affiliation{Niels Bohr Institute, University of Copenhagen, Lyngbyvej 2, DK-2100 Copenhagen $\O$, Denmark}
\affiliation{Cosmic DAWN Center, Denmark}

\author{Yasunori Terao}
\affiliation{Institute of Astronomy, Graduate School of Science, The University of Tokyo, 2-21-1 Osawa, Mitaka, Tokyo 181-0015, Japan}

\author[0000-0002-3801-434X]{Haruka Kusakabe}
\affiliation{Observatoire de Gen\`{e}ve, Universit\'e de Gen\`{e}ve, 51 chemin de P\'egase, 1290 Versoix, Switzerland}

\author{Makoto Ando}
\affiliation{Department of Astronomy, School of Science, The University of Tokyo, 
7-3-1 Hongo, Bunkyo-ku, Tokyo 113-0033, Japan}

\author{Kentaro Motohara}
\affiliation{National Astronomical Observatory of Japan, 2-21-1 Osawa, Mitaka, Tokyo 181-8588, Japan}
\affiliation{Institute of Astronomy, Graduate School of Science, The University of Tokyo, 2-21-1 Osawa, Mitaka, Tokyo 181-0015, Japan}

\author[0000-0001-5185-9876]{Lee Spitler}
\affiliation{Research Centre for Astronomy, Astrophysics $\&$ Astrophotonics, Macquarie University, Sydney, NSW 2109, Australia}
\affiliation{Department of Physics $\&$ Astronomy, Macquarie University, Sydney, NSW 2109, Australia}


\begin{abstract}

In order to shed light on how galactic properties depend on the intergalactic medium (IGM) environment traced by the Ly$\alpha$ forest, we observationally investigate the IGM--galaxy connection using the publicly available 3D IGM tomography data (CLAMATO) and several galaxy catalogs in the COSMOS field.
We measure the cross-correlation function (CCF) for $570$ galaxies with spec-$z$ measurements and detect a correlation with the IGM up to $50\,h^{-1}$\, comoving Mpc. 
We show that galaxies with stellar masses of $10^9-10^{10}$\,M$_\odot$ are the dominant contributor to the total CCF signal. 
We also investigate CCFs for several galaxy populations: Ly$\alpha$ emitters (LAEs), H$\alpha$ emitters (HAEs), [O\,{\sc iii}] emitters (O3Es), active galactic nuclei (AGNs), and submillimeter galaxies (SMGs), 
and we detect the highest signal in AGNs and SMGs at large scales ($r\geq5$~$h^{-1}$~Mpc), but in LAEs at small scales ($r<5$~$h^{-1}$~Mpc).
We find that they live in various IGM environments --
HAEs trace the IGM in a similar manner to the continuum-selected galaxies, but LAEs and O3Es tend to reside in higher-density regions. 
Additionally, LAEs' CCF is flat up to $r\sim3$\,$h^{-1}$\,Mpc, indicating that they tend to avoid the highest-density regions.
For AGNs and SMGs, the CCF peak at $r=5-6$\,$h^{-1}$\,Mpc implies that they tend to be in locally lower-density regions. 
We suspect that it is due to the photoionization of IGM {\sc Hi} by AGNs, i.e., the proximity effect.

\end{abstract}


\section{introduction}
The link between the intergalactic medium (IGM) and galaxies is key to understanding the evolution of baryonic matter and galaxies. This is because the IGM and galaxies continuously interact with each other --- galaxies are formed from condensed gas, increase their baryonic mass by accruing gas from the IGM, and pollute the surrounding IGM with metals. 

Observationally, the IGM gas can be probed by Ly$\alpha$ forest absorption in background quasars' (QSOs) and bright galaxies' spectra, which originates from the neutral atoms in photo-ionized gas (e.g., \citealp{miralda96,rauch98}). 
Its connection with galaxies has been investigated from the nearby universe to high redshift ($z=6$) in the literature
(e.g., \citealp{adelberger03,adelberger05,chen05,ryan-weber06,wilman07,FG08,chen09,rakic11,rakic12,rudie12,font-ribera13,profX13,tejos14,bielby17,meyer19b,meyer19a}). 

In particular, the IGM--galaxy connection has been examined by paying attention to overdense regions of galaxies
(e.g., \citealp{stark15,cai16,lee16}). 
\citet{cucciati14} have found a significant Ly$\alpha$ absorption feature at the redshift of a protocluster in a stacked spectrum of galaxies behind the protocluster.
\citet{mawatari17} have evaluated the IGM absorption enhancement with photometric images for the SSA22, Great Observatory Optic Deep Survey North (GOODS-N; \citealp{dicknson04}), and Subaru/XMM-Newton Deep Survey (SXDS; \citealp{furusawa08}) fields. They have found a clear enhancement of the IGM neutral hydrogen ({\sc Hi}) in the confirmed high galaxy density structure SSA22, but not in the remaining two fields.
Those studies have shown the presence of an IGM {\sc Hi} overdensity in cluster regions, and vice versa (e.g., \citealp{cai16,hayashino19}).
Using the spectra taken by the Baryon Oscillations Spectroscopic Survey project \citep{dawson13} of the Sloan Digital Sky Survey III \citep{einstein11}, \citet{cai16} have identified IGM overdense regions from the optical depth of IGM {\sc Hi}, and confirmed that those regions are also overdense in galaxies.
\citet{lee16} have found that an IGM overdensity region in their 3D tomography data of Ly$\alpha$ forest absorption coincides with a known protocluster at $z=2.45$ \citep{diener15,chiang15}.

The IGM--galaxy connection in the low-density environments of the field has been examined by cross-correlation between Ly$\alpha$ forest absorption and galaxies
(e.g., \citealp{adelberger05,font-ribera12,font-ribera13,tejos14,bielby17,mukae19}). 
Particularly, those studies have targeted specific galaxy populations, such as QSOs \citep{font-ribera13,profX13}, Lyman-break galaxies (LBGs) at $z=2-3$ (\citealp{adelberger03,adelberger05}; the Keck Baryonic Structure Survey, KBSS; e.g., \citealp{rakic11,rakic12,turner14}; the VLT LBG Redshift Survey, VLRS; e.g., \citealp{crighton11,tumm14,bielby17}), and damped Ly$\alpha$ systems (DLAs) at $z<1$ \citep{font-ribera12,rubin15,peres-rafols18,alonso18}. 
Those studies have detected a cross-correlation signal up to several tens of comoving $h^{-1}$\,Mpc scales. 

An alternative method for investigating the IGM--galaxy connection has been introduced by \citet{mukae17}, which enabled a comparison between the large-scale spatial distributions of galaxies and the IGM. They have compared IGM-overdensity $\delta_\text{F}$ and galaxy-overdensity ($\delta_\text{galaxy}$) evaluated from a cylinder of $\sim8.8$ comoving $h^{-1}$\,Mpc radius with $\sim88$ comoving $h^{-1}$ Mpc depth at $z\sim2.5$ and found an anti-correlation between these two parameters. They have suggested that the correlation is produced by filamentary large-scale structures of both the IGM and galaxies along the sightline.

Those previous studies have successfully identified the presence of IGM--galaxy connection, which continues to tens of comoving $h^{-1}$ Mpc scales. In addition, several studies have found some variation in the connection depending on galactic properties (e.g., \citealp{adelberger03,adelberger05,chen05,chen09,tejos14}).
However, the understanding of their variation over galaxy properties and populations is limited. 
In order to shed more light on the IGM--galaxy connection, we examine the connection between the IGM and galaxies using observational data.
We use the publicly available Ly$\alpha$ forest 3D tomography data of the COSMOS Ly$\alpha$ Mapping And Tomography Observations (CLAMATO; \citealp{lee14,lee16,lee18}) as the IGM gas and several galaxy catalogs in the literature.
Because we also compare the stellar mass and star formation rate (SFR) dependence of the IGM--galaxy cross-correlation function (CCF) obtained in this paper with that predicted from cosmological hydrodynamical simulations \citep{momose20a}, we adopt the same stellar mass and SFR binning as used in \citet{momose20a}.

Our paper consists of the following sections.
We introduce the data used in this study in Section 2, and the methodology in Section 3. Observational results are shown in Section 4. 
Discussion and implications indicated from our results are presented in Section 5. 
Finally, a summary is given in Section 6. 
Throughout this paper, we use a cosmological parameter set of ($\Omega_{\rm m}$, $\Omega_\Lambda$, $h$) = ($0.31$, $0.69$, $0.7$), which has been adopted in the CLAMATO data \citep{lee16,lee18}. All distances are comoving, unless otherwise stated. 
In this paper, ``cosmic web'' and ``IGM'' indicate those traced by neutral {\sc Hi} gas unless otherwise specified.


\section{Data}

\subsection{The IGM Data}
\label{sec:igmdata}

We use the CLAMATO data as a tracer of IGM {\sc Hi} gas \citep{lee16,lee18}\footnote{The data is from: http://clamato.lbl.gov}. 
The CLAMATO is a 3D tomography map of $\delta_\text{F}$ over $2.05<z<2.55$ in $0.157$ deg$^2$ of the COSMOS field \citep{scovil07}.
Here, $\delta_\text{F}$ is the Ly$\alpha$ forest transmission fluctuation defined by 
\begin{equation}
    \delta_\text{F}=\frac{F}{\langle F_z \rangle} - 1,
\end{equation}
where $F$ and $\langle F_z\rangle$ are the Ly$\alpha$ forest transmission and its cosmic mean.
\citet{lee18} have measured $F$ using spectra of $240$ galaxies and QSOs taken with the LRIS spectrograph \citep{oke95,steidel04} on Keck I. Those $240$ background objects are at $2.17<z<3.00$ and have [$2.61$, $3.18$] $h^{-1}$ Mpc separations on average at $z=2.3$ in [R.A., decl.] directions. 
The effective transverse separation is $2.04$ $h^{-1}$ Mpc.
The separation in the line-of-sight direction is $2.35$ $h^{-1}$ Mpc at $z=2.3$.
\citet{lee16,lee18} have evaluated $\delta_\text{F}$ with these spatial resolutions using $\langle F_z\rangle$ presented by \citet{FG08} and then reconstructed $\delta_\text{F}$ with the Wiener filtering algorithm to produce a 3D tomographic map. The final 3D data cube of the CLAMATO spans comoving dimensions of $(x, y, z) = (30, 24, 438)$ $h^{-1}$ Mpc, with a pixel size of $0.5$ $h^{-1}$ Mpc.



\begin{figure*}
	\begin{center}
	\includegraphics[width=\linewidth]{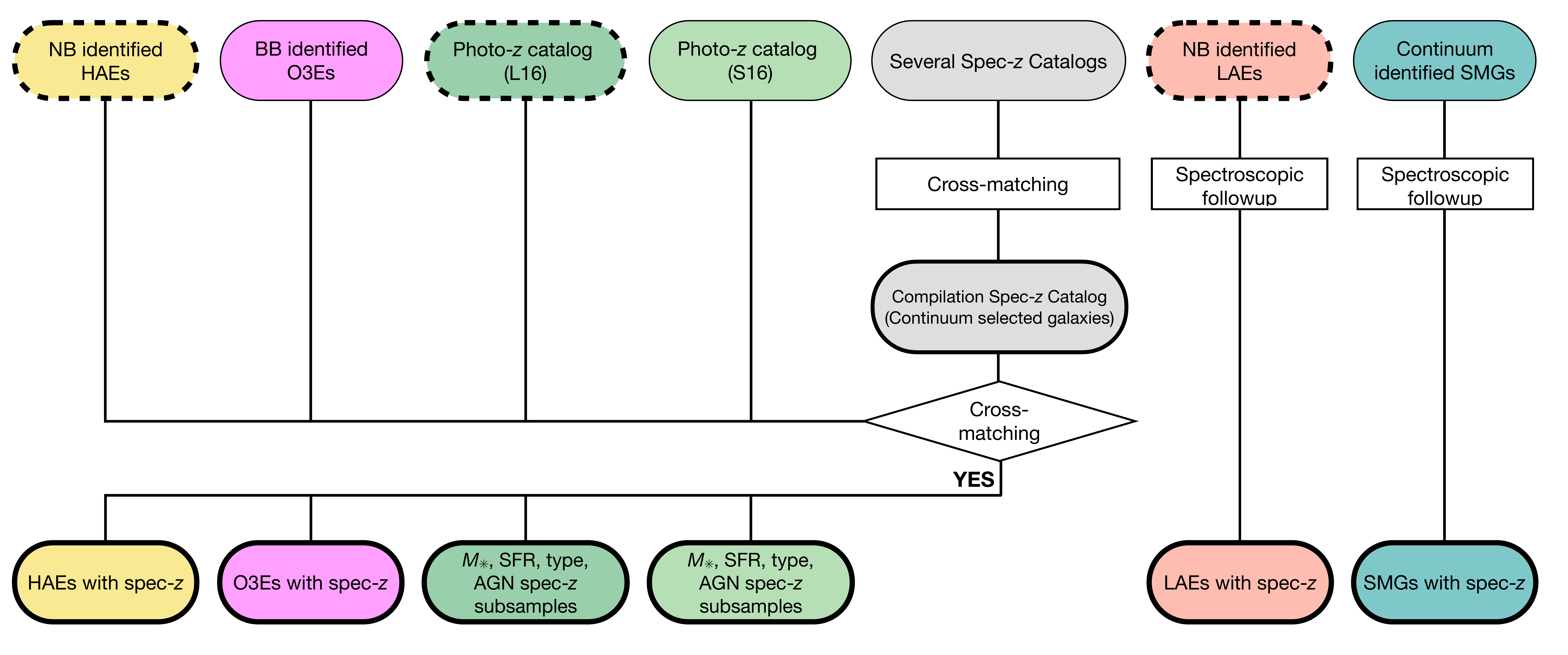} 
	\caption{
	Flowchart of catalog constructions. Each color indicates a different galaxy population. Edges indicated by thick solid and thick dashed lines represent catalogs used in cross-correlation and overdensity analyses, respectively.
	}
	\label{fig:flowchart}
	\end{center}
\end{figure*}

\begin{figure}
	\begin{center}
	\includegraphics[width=\linewidth]{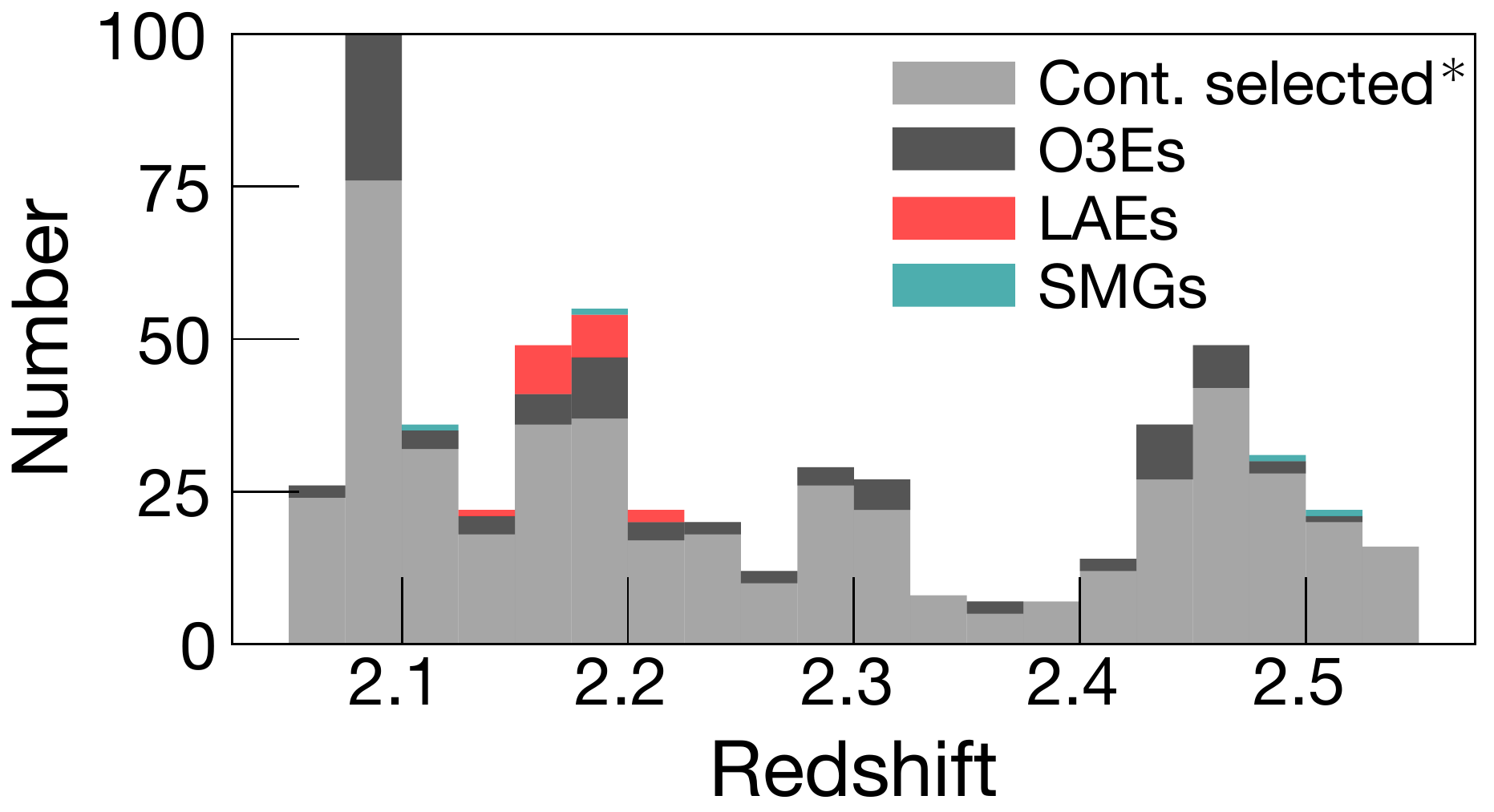} 
	\caption{
    Stacked number histogram of our sample galaxies used in cross-correlation analysis as a function of redshift. Note that O3Es are included in continuum-selected galaxies. Thus, the actual number of continuum-selected galaxies is the sum of Cont. selected* (light gray) and O3Es (dark gray).
	}
	\label{fig:hist_z}
	\end{center}
\end{figure}



\begin{table*}
    \caption{Number of Galaxies Used in Cross-correlation Analysis}
\begin{center}
\begin{tabular}{rrrrrrr}
\hline
    Continuum-selected & L16$^a$ & S16$^a$ & LAEs & HAEs & O3Es & SMGs  \\
\hline
    570 & 305 & 410 & 19   & 7    & 85 & 4     \\
\hline
\end{tabular}
\label{tab:num_sum1}
	\begin{tablenotes}
		\small
		$^a$ Number of cross-matched galaxies with the compiled spec-$z$ catalog.
    \end{tablenotes}
\end{center}
\end{table*}

\begin{table*}
    \caption{Subsamples of the L16 and S16 Spec-$z$ Samples Used for Overdensity Analysis}
\begin{center}
\begin{tabular}{llrrl}
\hline
    Category & Range/Subsample & $N_\text{L16}$ & $N_\text{S16}$ & Sample Name \\
\hline
    Stellar mass [$M_\sun$] & $\Mstar\geq10^{11}$ & $11$ & $14$  & $\Mstar$--$11$ \\
                            & $10^{10}\leq \Mstar <10^{11}$ & $107$    & $158$ & $\Mstar$--$10$\\
                            & $10^{9}\leq \Mstar < 10^{10}$  & $170$    & $210$ & $\Mstar$--$9$\\
                            & $\Mstar<10^{9}$ & $9$      & $7$   & $\Mstar$--$8$\\
\hline
    $\log$ (SFR/$M_\sun$ yr$^{-1}$)
        & $2\leq$ $\log$ SFR & $23$  & $5$   & SFR--(i) \\
        & $1\leq$ $\log$ SFR $<2$ & $207$ & $113$ & SFR--(ii) \\
        & $0\leq$ $\log$ SFR $<1$ & $56$  & $210$ & SFR--(iii)\\
        & $0>$ $\log$ SFR   & $11$  & $60$  & SFR--(iv)\\
\hline
    $\log$ (sSFR/yr$^{-1}$)
        & $-9\leq$ $\log$ sSFR & $245$ & $138$ & sSFR--(i) \\
        & $-10\leq$ $\log$ sSFR $<-9$    & $39$  & $191$ & sSFR--(ii) \\
        & $-10>$ $\log$ sSFR   & $11$  & $59$  & sSFR--(iii) \\
\hline
    AGNs    & Total            & $8$ & $21$ & AGN    \\
            & X-ray identified & $8$ & $16$ & - \\
            & IR identified    & $0$ & $8$  & - \\
\hline
    Galaxy type & Star-forming & $284$ & $389$ & SFG    \\
                & Quiescent    & $13$  &  --   & QG     \\
\hline
\end{tabular}
\label{tab:num_photoz}
\end{center}
\end{table*}


\subsection{Galaxy Samples}
\label{sec:catalogdata}

We use several spec-$z$ catalogs \citep{lilly07,lilly09,trump09,balo14,le15,kriek15,nana16,mom16,vanderwel16,mast17,hasi18}, two photo-$z$ catalogs (\citealp{laigle16,straa16}: hereafter L16 and S16, respectively), and catalogs of Ly$\alpha$ emitters (LAEs) at $z=2.14-2.22$ \citep{naka12,nakajima13,hashimoto13,shibuya14b, konno16}, H$\alpha$ emitters (HAEs) at $z=2.215-2.247$ \citep{sob13}, [{\sc Oiii}]$\lambda\lambda4959$, $5007$ emitters (O3Es) at $z=1.95-2.55$ (Y. Terao et al., in preparation), and submillimeter galaxies (SMGs) with spec-$z$ measurements \citep{smol12,bris17,micha17}. 
Galaxies with spec-$z$ measurements are used in the cross-correlation analysis (see also Section \ref{sec:intro_IM}), while those with photo-$z$ estimates alone and line emitters with and without spectroscopic redshifts are used in overdensity analysis. A detailed description is given in Section \ref{sec:intro_overden}.
Figure \ref{fig:flowchart} summarizes the catalogs used in this study, together with the catalog construction methodology.
We show the redshift distributions of our samples used in the cross-correlation analysis in Figure \ref{fig:hist_z}. Note that HAEs, O3Es, and active galactic nuclei (AGNs) are included in the continuum-selected galaxies in the compilation spec-$z$ catalog.
The following is a detailed description of the catalogs. The number of galaxies is summarized in Tables \ref{tab:num_sum1} and \ref{tab:num_photoz}.


\subsubsection{Continuum-selected Galaxies}
\label{sec:catalog_generalgal}

The cross-correlation analysis needs a spec-$z$ catalog (e.g., \citealp{momose20a}).
First, we compile all available spec-$z$ catalogs in the archive \citep{lilly07,lilly09,trump09,balo14,le15,kriek15,mom16,nana16,vanderwel16,mast17,hasi18} and construct one spec-$z$ catalog. 
We cross-match two catalogs with a maximum allowable separation of $1\arcsec$. If a galaxy is found in two or more catalogs, the spec-$z$ measurement obtained from near-IR observations or with a better quality flag in an original catalog is adopted. The final cross-matched spec-$z$ catalog consists of $570$ galaxies. Hereafter we refer to the catalog and galaxies in it as ``compiled spec-$z$ catalog'' and ``continuum-selected galaxies''.

For each galaxy in the compiled spec-$z$ catalog, we take stellar mass ($\Mstar$), SFR, and specific SFR (sSFR) estimates, AGN flag, and galaxy type flag (either star-forming or quiescent) from the existing photo-$z$ catalogs of L16 and S16.
Since $\Mstar$ and SFR are obtained by spectral energy distribution (SED) fitting based on photometric images and depend on the set of galaxy models, we use two independent photo-$z$ catalogs of L16 and S16.
Note that the survey field of L16 covers the entire CLAMATO field, while that of S16 is included in the CLAMATO field and covers only $24\%$ of it (see also Fig. 1 in L16, Fig. 7 in S16, and Fig. 1 in \citealp{lee18}). 
L16 have used {\sc LePhare} to compute photometric redshifts \citep{arnouts02,ilbert06}, while S16 have used EASY \citep{brammer08}. Both studies have calculated photo-$z$ and SED with near ultraviolet (NUV), optical, near-infrared (NIR), and mid-IR (MIR) data. Note that, although with a smaller survey field, S16 have used deeper NIR images and thus obtained better photo-$z$ accuracy than L16.
The photo-$z$ uncertainties of L16 and S16 are $\sigma_{\Delta z}$/$(1+z)=0.021$ and $0.009$ with catastrophic errors of $13.2$~$\%$ (at $3<z<6$) and $2.4$~$\%$, respectively.
Both L16 and S16 have assumed a \citet{chabrier03} initial mass function. 
A galaxy type either star-forming or quiescent in L16 has been determined from the color-color diagram of the NUV $-$ $r$/$r - J$ (see more detail in L16).
We perform cross-matching between the compiled spec-$z$ catalog and the photo-$z$ catalogs with a radius of $0\farcs5$; $305$ and $410$ galaxies in the compiled spec-$z$ catalog are cross-matched with L16 and S16, respectively. 
We should note that owing to the deeper limiting magnitude in $K_\text{S}$ band of S16, they have detected about a factor of $4$ more objects than L16 (see Fig.13 of S16 and Fig.7 of L16). Thus, the numbers of galaxies of L16 and S16 used in this study are similar, though S16 only cover $24\%$ of the CLAMATO field.

There are $4715$ ($1934$) photo-$z$ galaxies in L16 (S16) within the CLAMATO field. Nevertheless, due to large photo-$z$ errors with $\sigma_z = 0.07$ for L16 ($0.023$ for S16) corresponding to $61$ ($21$) $h^{-1}$ Mpc at $z=2.3$, we only use them for overdensity analysis.


\subsubsection{AGNs}
\label{sec:catalog_AGN}

We construct AGN catalogs by cross-matching L16 and S16 with the compiled spec-$z$ catalog. Finally, we have 8 and 21 AGNs from L16 and S16, respectively. 
For L16-AGNs, we regard a galaxy with X-ray flag as an AGN, meaning that L16-AGNs are X-ray-identified AGNs. 
For S16-AGNs, we select galaxies with AGN flags given in S16.
Because S16-AGNs have been identified by IR, radio, and X-ray emission \citep{Cowley16}, we also use this information in the CCF analysis. Among the $21$ S16-AGNs, ($8$, $1$, $21$) are identified in (IR, radio, X-ray), where four are classified as both IR and X-ray AGNs.
Note that four AGNs are common to L16 and S16.


\begin{figure*}
	\begin{center}
	\includegraphics[width=\linewidth]{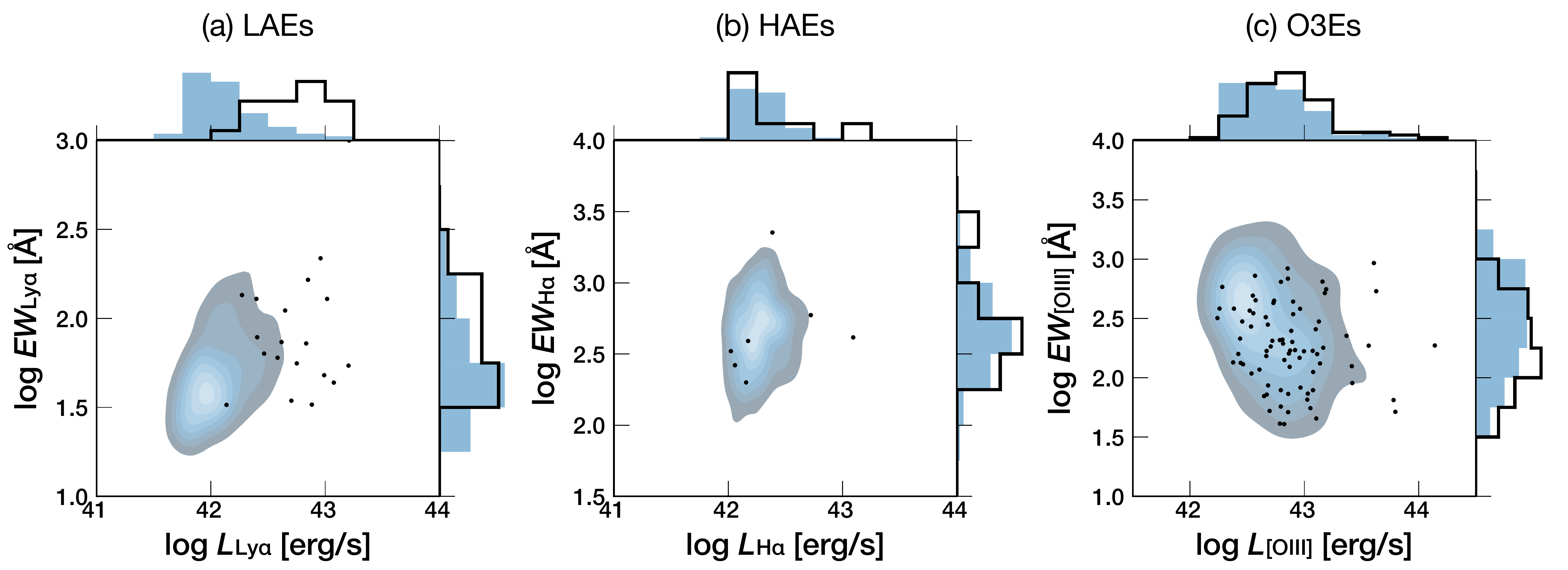} 
	\caption{
    Distribution of the luminosity and {\it EW} of (a) LAEs, (b) HAEs, and (c) O3Es with and without spec-$z$ measurements. Contours are for all emitters, while black dots are for those with spec-$z$ measurements. Filled and open histograms indicate all and spec-$z$ emitters, respectively.
	}
	\label{fig:prop_emitter}
	\end{center}
\end{figure*}


\subsubsection{Line Emitters}
\label{sec:catalog_emitter}

There are ($358$, $44$, $575$) photometrically identified (LAEs, HAEs, O3Es) in the CLAMATO field (LAEs: \citealp{naka12,konno16}, HAEs: \citealp{sobral13}, O3Es: Y. Terao et al., in preparation).
Note that O3Es have been identified in the S16 field that is smaller than that of CLAMATO. 
Among them, all LAEs and HAEs are used for overdensity analysis; note that we do not use O3Es for the analysis, because they have been found by an excess of a broad-band filter, which covers a much wider redshift range with $1.95<z<2.55$.
For cross-correlation analysis, we only use those with spec-$z$ measurements at $2.05<z<2.55$.

Spec-$z$ measurements of LAEs are taken from \citet{hashimoto13}, \citet{nakajima13}, and \citet{shibuya14b}. Among the $358$ narrow-band ($NB$) identified LAEs, only $19$ have spec-$z$ measurements. We should note that we only use LAEs whose redshifts are determined not by Ly$\alpha$ line but by nebular lines (e.g., H$\alpha$ and [{\sc Oiii}]) for the cross-correlation analysis, because the redshift by Ly$\alpha$ line is known to be larger by more than $100$ km s$^{-1}$ than the systemic redshift measured by nebular lines (e.g., \citealp{fink11,Mclinden11,hashimoto13,hashimoto15,erb14,shibuya14b,song14}). 

Unfortunately, no spectroscopic redshifts are given in the original HAE and O3E catalogs. Thus, we conduct cross-matching of their catalogs with the compiled spec-$z$ catalog using a searching radius of $0\farcs5$, which is the same value as used for cross-matching with the photo-$z$ catalogs. 
Among the $44$ HAEs ($575$ O3Es), $7$ ($85$) have spec-$z$ measurements.
The redshifts of two HAEs among the seven are not in the range expected from the full-width half maximum (FWHM) of the $NB$ filter, but still in the range where the filter has a sensitivity. In addition, no galaxies are found within $2\arcsec$ radius around them. Hence, we include those two HAEs for cross-correlation analysis. 
We should also note that two O3Es each have two counterparts in the compiled spec-$z$ catalog. We adopt the redshift of the galaxy that is closer to the position of the O3E. The contribution by those two O3Es to our CCFs is, however, negligible. 

Our emitter samples, particularly those with spec-$z$ measurements, may be dual or triple emitters. For instance, all of our LAEs with spec-$z$ measurements have H$\alpha$ and/or [O{\sc iii}] detections and thus can be also regarded as HAEs and/or O3Es.
However, in this study, we classify emitters based on their first identification by photometric images owing to the lack of luminosity or equivalent width (\textit{EW}) measurements by other emission lines. 
For example, LAEs with spec-$z$ measurements are not included in either the HAE or O3E sample.

In order to assess whether line-emitting galaxies with spec-$z$ measurements represent their parent sample, we compare luminosity ($L_\text{Ly$\alpha$}$, $L_\text{H$\alpha$}$, and $L_\text{[{\sc Oiii}]$\lambda\lambda5007$}$) and {\it EW} ($EW_\text{Ly$\alpha$}$, $EW_\text{H$\alpha$}$, and $EW_\text{[{\sc Oiii}]$\lambda\lambda5007$}$) between the parent and spec-$z$ samples in Figure \ref{fig:prop_emitter}. 
We find LAEs with spec-$z$ measurements to be biased toward high Ly$\alpha$ luminosities. The difference in the $L_\text{Ly$\alpha$}$ distribution is also confirmed by the Kolmogorov-Smirnov (KS) test, which gives $p$-values of $9.4\times10^{-9}$ and $0.02$ for the $L_\text{Ly$\alpha$}$ and $EW_\text{Ly$\alpha$}$ distributions, respectively.
Similarly, O3Es with spec-$z$ measurements are biased toward higher [{\sc Oiii}] $\lambda\lambda5007$ luminosities with $p$-values of $9.6\times10^{-3}$ and $0.03$ for the $L_\text{[{\sc Oiii}]$\lambda\lambda5007$}$ and $EW_\text{[{\sc Oiii}]$\lambda\lambda5007$}$ distributions.
For HAEs, on the other hand, we do not find a clear difference in either the luminosity or {\it EW} distribution.


\subsubsection{SMGs}
\label{sec:catalog_SMG}

We find $24$ SMGs in the spec-$z$ catalogs within the CLAMATO volume \citep{smol12,bris17,micha17}. However, most of them have a relatively large spec-$z$ error ($\sigma_z >0.1$). 
Therefore, we only use four SMGs whose redshifts have been measured by NIR or optical spectroscopy with a sufficiently small error ($\sigma_z \leq0.0023$ corresponding to $2$ $h^{-1}$ Mpc at $z=2.3$).


\section{Methodology}

In order to investigate the connection between the IGM and galaxies, we apply two methods -- we refer to them as ``cross-correlation analysis'' and ``overdensity analysis''. Each method is introduced in the following subsections in detail.

\subsection{Cross-Correlation Analysis}
\label{sec:intro_IM}

The first method is the cross-correlation between CLAMATO and galaxies with spec-$z$ measurements. The CCF used in this study is  
\begin{equation}
\begin{split}
  \xi_\text{$\delta$F}(r) 
  &= \frac{1}{\sum_{i=1}^{N(r)} \omega_{g, i}} \sum_{i=1}^{N(r)} \omega_{g, i} \delta_{g, i} \\
  &- \frac{1}{\sum_{j=1}^{M(r)} \omega_{ran, j}} \sum_{j=1}^{M(r)} \omega_{ran, j} \delta_{ran, j}
  \label{eq:cross-corr}
\end{split}
\end{equation}
\begin{equation}
    \omega_{g, i} = \frac{1}{(\sigma_{g, i})^2}, \  
    \omega_{ran, j} = \frac{1}{(\sigma_{ran, j})^2}
    \label{eq:cross-corr-sigma}
\end{equation}
where $\xi_\text{$\delta$F}$ is the cross-correlation at a separation $r$; 
$\delta_{g, i}$ ($\delta_{ran, j}$) and $\sigma_{g, i}$ ($\sigma_{ran, j}$) are the Ly$\alpha$ forest transmission fluctuation at a place $i$ ($j$) separated by $r$ from a galaxy (random point) and its error, respectively. 
Here, $N(r)$ and $M(r)$ represent the numbers of pixel-galaxy and pixel-random pairs with separation $r$, respectively. 
We adopt the CLAMATO's 3D noise standard deviation measurements $\sigma$ as $\sigma_{g}(r)$ and $\sigma_{ran}(r)$. The CLAMATO's standard deviation cube includes pixel noise, finite skewer sampling, and the intrinsic variance of the Ly$\alpha$ forest (see details in \citealp{lee18}).
Note that $r$ used for the cross-correlation analysis is 3D radius.
This method is often adopted to measure the large-scale Ly$\alpha$ intensity (e.g., \citealp{croft16,croft18,kakuma19}). 
We calculate $\xi_\text{$\delta$F}$ for a series of spherical shells from $r=1.3$ to $100$ ($10^{0.1}$ to $10^2$) $h^{-1}$ Mpc with a $\Delta \log (r/h^{-1}$ Mpc) $=0.1$ interval. 

The statistical errors in the CCF are evaluated using the jackknife resampling method. Usually, jackknife resampling for the spatial cross-correlation is performed by dividing the survey volume into several small subvolumes and removing one at a time.
However, this usual method gives extremely small errors, because our galaxy samples are small and their sky distribution is biased toward regions where intensive spec-$z$ follow-up observations have been conducted. Instead, we perform resampling by removing one object from the given sample and calculating a CCF, and by repeating this process for the number of objects in the sample. This means that the number of jackknife samples in each sample presented in Tables~\ref{tab:num_sum1} and \ref{tab:num_photoz} is the same as that of galaxies in it. Because of that, the CCF errors in this study are dominated by the small sample sizes.

\begin{table*}
    \caption{Measurement results of overdensity analysis from local minima and maxima.}
\begin{center}
\begin{tabular}{crrrrrr}
\hline
    Sample & $N_1^{(1)}$ & $N_2^{(2)}$ & $R_s^{(3)}$ & $p^{(3)}$ & $\alpha^{(4)}$ & $\beta^{(4)}$\\
\hline \hline
    L16-$\Mstar$--$10$ & 137 & 41 & $-0.31$ & $0.30$ & $0.003\pm0.011$ & $-0.061\pm0.023$  \\
    L16-$\Mstar$--$9$  & 462 & 159 & $-0.54$ & $0.05$ & $-0.006\pm0.010$ & $-0.121\pm0.040$ \\ 
    L16-$\Mstar$--$8$  & 266 & 86 & $-0.17$ & $0.58$ & $-0.009\pm0.012$ & $-0.016\pm0.034$ \\
\hline
    ALL & 865 & 286 & $-0.47$ & $0.10$ & $-0.006\pm0.011$ & $-0.085\pm0.040$ \\
\hline
    LAEs & 358 & -- & $-0.70$ & $0.19$ & $-0.088\pm0.021$ & $-0.147\pm0.054$ \\
\hline
    HAEs & -- & 44 & $-0.19$ & $0.65$ & $-0.002\pm0.014$ & $-0.010\pm0.023$ \\
\hline
\end{tabular}
\label{tab:overdensity}
\end{center}
	\begin{tablenotes}
		\small
		\item $^{(1)}$ Number of galaxies in $2.14<z<2.22$. $^{(2)}$ Number of galaxies in $2.215<z<2.247$.
		$^{(3)}$ Spearman's coefficient and $p-$value for the $\langle \delta_\text{F} \rangle - \delta_\text{galaxy}$ relation from local minima and maxima. 
		$^{(4)}$ The best-fit parameters of chi-square fitting of the $\langle \delta_\text{F} \rangle - \delta_\text{galaxy}$ relation from local minima and maxima. 
	\end{tablenotes}
\end{table*}

\begin{table*}
    \caption{Measurement results of overdensity analysis from random positions.}
\begin{center}
\begin{tabular}{crrrrrr}
\hline
    Sample & $N_1^{(1)}$ & $N_2^{(2)}$ & $R_s^{(3)}$ & $p^{(3)}$ & $\alpha^{(4)}$ & $\beta^{(4)}$\\
\hline \hline
    L16-$\Mstar$--$10$ & 137 & 41 & $-0.79$ & 3.75e-3 & $-0.006\pm0.007$ & $-0.057\pm0.011$ \\
    L16-$\Mstar$--$9$  & 462 & 159 & $-0.65$ & $0.02$ & $-0.014\pm0.007$ & $-0.156\pm0.025$ \\
    L16-$\Mstar$--$8$  & 266 & 86 & $0.21$ & $0.49$ & $-0.004\pm0.026$ & $0.424\pm0.237$ \\
\hline
    ALL & 865 & 286 & $-0.65$ & $0.02$ & $-0.011\pm0.007$ & $-0.170\pm0.029$ \\
\hline
    LAEs & 358 & -- & $-0.70$ & $0.19$ & $-0.054\pm0.011$ & $-0.059\pm0.043$  \\
\hline
    HAEs & -- & 44 & $-0.26$ & $0.53$ & $-0.003\pm0.018$ & $-0.038\pm0.037$   \\
\hline
\end{tabular}
\label{tab:overdensity_ran}
\end{center}
	\begin{tablenotes}
		\small
		\item $^{(1)}$ Number of galaxies in $2.14<z<2.22$. $^{(2)}$ Number of galaxies in $2.215<z<2.247$.
		$^{(3)}$ Spearman's coefficient and $p-$value for the $\langle \delta_\text{F} \rangle - \delta_\text{galaxy}$ relation from random points. 
		$^{(4)}$ The best-fit parameters of chi-square fitting of the $\langle \delta_\text{F} \rangle - \delta_\text{galaxy}$ relation from random points. 		
	\end{tablenotes}
\end{table*}


\subsection{Overdensity Analysis}
\label{sec:intro_overden}

Because our samples of galaxies with spec-$z$ measurements are very limited, we also apply another analysis to use as many galaxies as possible. 
This second method compares mean IGM fluctuations ($\langle\delta_\text{F}\rangle$) and galaxy overdensities within randomly distributed cylinders \citep{mukae17}, and can be applied to photometric redshift samples whose redshift uncertainties $\sigma_z$ are less than $0.1$. 
In this study, we focus only on two redshift ranges of $2.14<z<2.22$ and $2.215<z<2.247$, which are defined by the central wavelength and FWHM of the $NB$ filters for LAEs and HAEs. Additionally, we only use photo-$z$ galaxies of L16 for the overdensity analysis, because the survey area of S16 is smaller than CLAMATO's coverage. The number of galaxies used in the analysis is shown in Tables \ref{tab:overdensity} or \ref{tab:overdensity_ran}.

We first collapse each of the CLAMATO data of the above two redshift ranges in the redshift direction to generate a 2D map, where the thickness of the original data, $\Delta z=0.08$ ($0.032$), corresponds to $69.7$ ($27.9$) $h^{-1}$ Mpc for LAEs (HAEs).
Then, for each 2D map, we identify local minima and maxima of the IGM fluctuations and calculate $\langle \delta_\text{F} \rangle$ within a circle of radius $r$ centered at them. 
Since the original data have a thickness of $\Delta z$, $\langle \delta_\text{F} \rangle$ effectively means the mean IGM fluctuations within a cylinder whose volume is 
$\Delta z \times \pi r^2$ and is calculated with
\begin{equation}
  \langle \delta_\text{F} \rangle = \frac{1}{\sum_{i=1}^{N(r)} \omega_{g, i}(r)} \sum_{i=1}^{N(r)} \omega_{g, i}(r) \delta_{g, i}(r),
\end{equation}
where $\delta_{g,i}$ and $\omega_{g,i}$ are the same as in Equations \ref{eq:cross-corr} and \ref{eq:cross-corr-sigma} but obtained from the 2D CLAMATO map. We should note that we also generate 2D standard deviation maps in order to evaluate $\omega_{g,i}$ from the 2D map.
For a direct comparison between the above two redshift ranges, we calculate $\langle \delta_\text{F} \rangle$ with the same volume by adopting a different cylinder radius, that is, $3$ and $4.74$ $h^{-1}$ Mpc for $2.14<z<2.22$ and $2.215<z<2.247$, respectively. 
The radii are determined to satisfy three requirements:
i) they are larger than the transverse resolution of the CLAMATO data (see Section \ref{sec:igmdata}),  
ii) they are sufficiently small so that neighboring cylinders do not significantly overlap with each other, and
iii) requirements i) and ii) are satisfied in both two redshift ranges. 

Galaxy overdensities are evaluated within the same cylinders. 
We calculate galaxy overdensity ($\Sigma_\text{gal}$) with
\begin{equation}
    \Sigma_\text{gal} = \frac{N_\text{gal}}{\langle N_\text{gal} \rangle} - 1,
\end{equation}
where $N_\text{gal}$ is the number of galaxies in the cylinder and $\langle N_\text{gal} \rangle$ is the mean number of galaxies expected to be found in the same volume. 
We estimate the error of $\Sigma_\text{gal}$ with the Poisson errors.
Note that we do not consider any photo-$z$ uncertainties including catastrophic errors here, because they influence both $N_\text{gal}$ and $\langle N_\text{gal} \rangle$ estimates. Qualitatively, those errors will weaken the contrast of galaxy over/underdensities, which results in a narrower dynamic range of $\Sigma_\text{gal}$.

In order to examine whether the measured correlation is biased owing to using only local maxima and minima positions, we also investigate the $\langle\delta_\text{F}\rangle$--$\Sigma_\text{gal}$ relation based on randomly chosen cylinder positions (see also \citealp{mukae17}). If photo-$z$ measurements are valid with smaller errors than the cylinder depth, the bias should be negligible (\citealp{mukae17,momose20a}).


\section{Results}


\begin{figure}
	\begin{center}
	\plotone{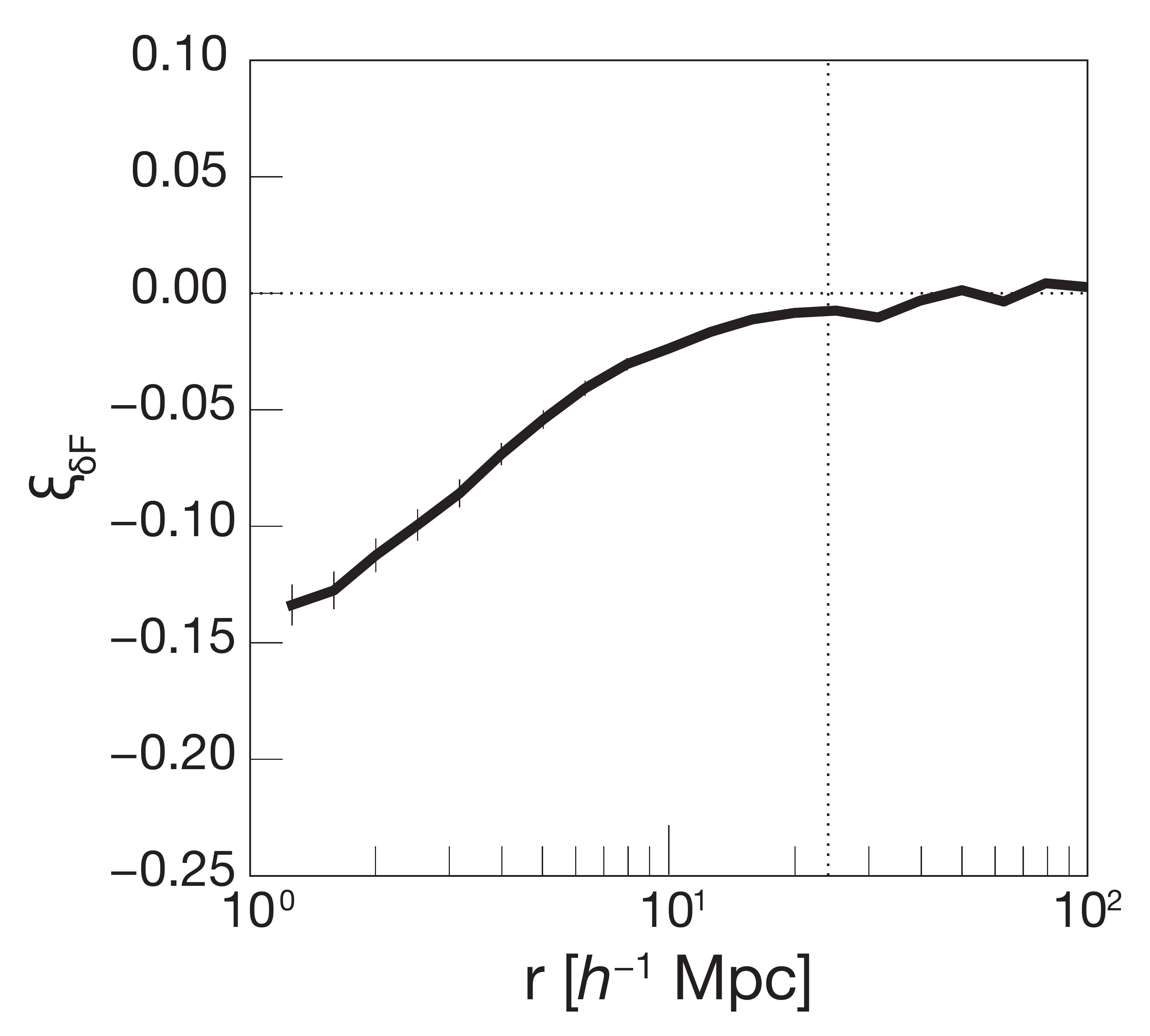}
	\caption{
	The CCF of all galaxies in our compiled spec-$z$ catalog.
	A vertical dotted line shows the possible largest radius for 3D cross-correlation calculations. Beyond this radius, the majority of $\delta_\text{F}$ values are from the line-of-sight direction.}
	\label{fig:IM_specall}
	\end{center}
\end{figure}


\subsection{Cross-correlation Analysis}
\label{sec:result_IM}

In this section, we show results of the cross-correlation analysis. 
Figure \ref{fig:IM_specall} shows the CCF from all galaxies in the compiled spec-$z$ catalog. 
A strong signal is detected at the center with $\xi_\text{$\delta$F}=-0.14$. The CCF increases monotonically  and reaches the cosmic mean ($\xi_\text{$\delta$F}=0$) at $r\sim50$ Mpc $h^{-1}$. 
If the IGM {\sc Hi} density around galaxies is higher than the mean {\sc Hi} density, the CCF has a negative value because of stronger Ly$\alpha$ absorption. Thus, Figure \ref{fig:IM_specall} indicates that galaxies are in {\sc Hi} overdensity regions on average up to $\sim50$ $h^{-1}$ Mpc in radius.
We should note that we cannot calculate a CCF three-dimensionally over $r>12$ $h^{-1}$ Mpc because of the limitation of the CLAMATO volume. 


\begin{figure*}
	\begin{center}
	\includegraphics[width=\linewidth]{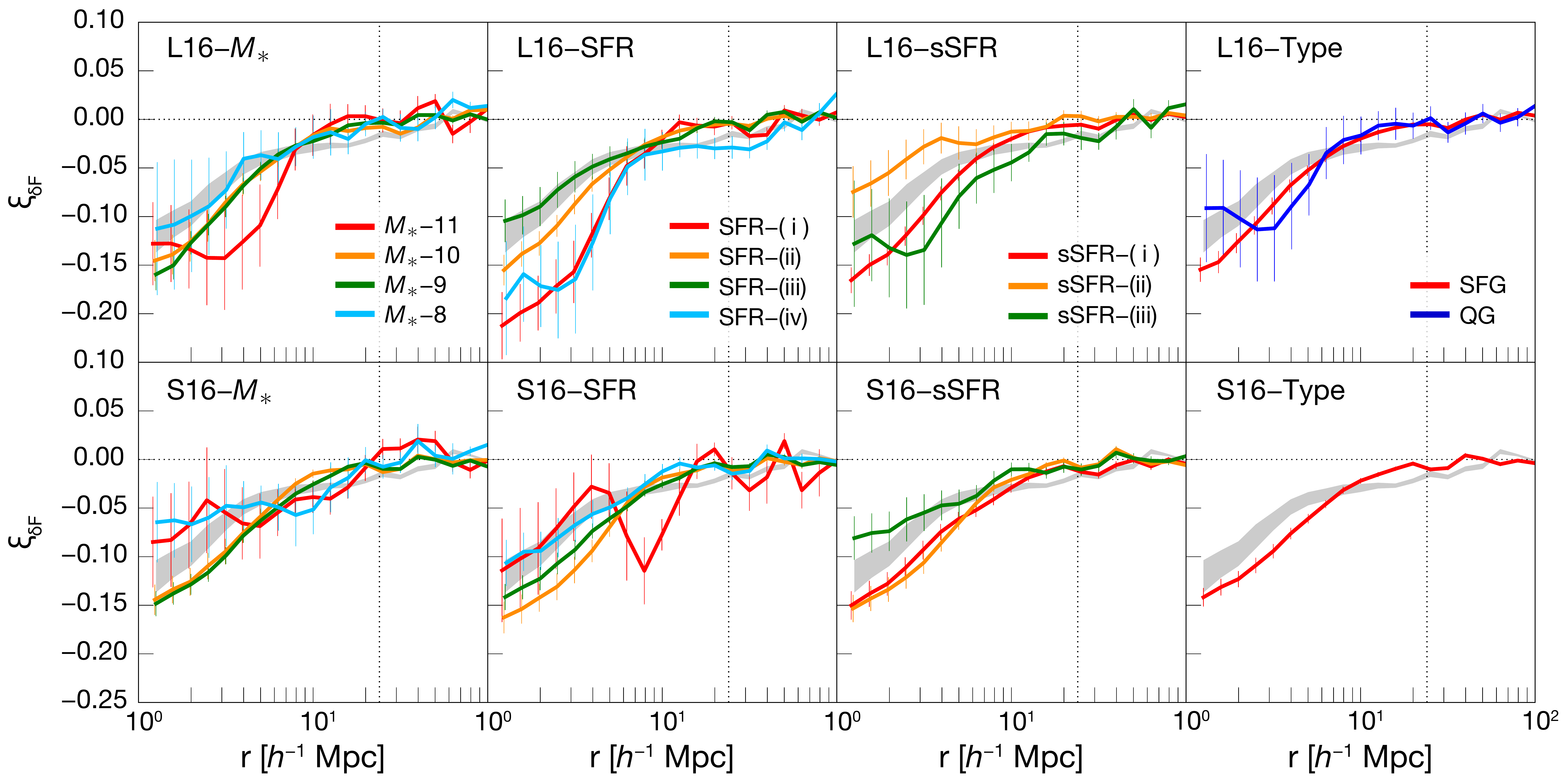} 
	\caption{
	CCFs of (top) L16 and (bottom) S16 subsamples divided by
	$\Mstar$, SFR, sSFR, and galaxy type (from left to right) as a function of radius in comoving units.
    The meaning of vertical dotted lines is the same as in Figure \ref{fig:IM_specall}.
    Gray shades represent the CCF of continuum-selected galaxies in the compiled spec-$z$ catalog.
    }
	\label{fig:IM_obscomp1}
	\end{center}
\end{figure*}


\subsubsection{Galaxy Properties}
\label{sec:result_IM_galpro}

For more detailed investigations depending on galaxy properties, we divide the L16 and S16 samples into four or three subsamples based on $\Mstar$, SFR, sSFR, and galaxy type (star-forming or quiescent galaxy, hereafter SFG or QG). 
For a direct comparison with the results from hydrodynamical simulations given in \citet{momose20a}, we set the subsample ranges to be basically the same as those used in \citet{momose20a}.
The number of galaxies in each subsample and its sample name are listed in Table \ref{tab:num_photoz}. 
We show the CCFs of individual samples in Figure \ref{fig:IM_obscomp1}. Because we take $\Mstar$, SFR, and sSFR values from both L16 and S16, we regard a sample for which L16 and S16 give consistent CCFs as being reliable. 

For all the subsamples, we detect a signal up to $r=10-20$ $h^{-1}$ Mpc. Nevertheless, we do not find any trends depending on $\Mstar$, SFR, or sSFR for either L16 or S16. In addition, due to the large error bars in several subsamples ($\Mstar$--$11$, $\Mstar$--$8$, SFR--(i), SFR--(iv), L16-sSFR--(iii), and L16-QG), the variation of the CCF depending on mass, SFR, and sSFR that has been confirmed in the simulations \citep{momose20a} is insignificant.
Detailed discussion on the lack of significant dependence on galactic properties is given in Section~\ref{sec:dis_absent_trend}.
Note that photo-$z$ errors do not affect the CCF calculation itself, because the calculation only uses spec-$z$ measurements as the line-of-sight positions of galaxies. However, the photo-$z$ errors affect the grouping depending on galactic properties as a contamination to subsamples. This effect seems to be larger for fainter galaxies (e.g., Fig. 21 in S16), and attenuates the dependence of the CCFs on galactic properties even if it exists \citep{momose20a}.

For a comparison between L16 and S16, several subsamples show consistency. 
Among the $\Mstar$ subsamples, the $\Mstar$--$10$ and $\Mstar$--$9$ subsamples give consistent CCFs with the strongest signal at the center, $\xi_\text{$\delta$F}=-0.15$, and reaching the cosmic mean at $r=20-30$ $h^{-1}$ Mpc. 
For the SFR subsamples, only the SFR--(ii) subsamples give consistent results, showing a monotonic increase with the strongest signal at the center of $\xi_\text{$\delta$F}=-0.16$. 
For the sSFR subsamples, only the sSFR--(i) subsamples give consistent results, with the strongest signal of $\xi_\text{$\delta$F}=-0.15 - -0.17$.
The SFG subsamples of L16 and S16 show similar monotonically increasing CCFs starting from $\xi_\text{$\delta$F}=-0.15$.

In order to quantify the similarity or difference in the CCF between L16 and S16 subsamples and all continuum-selected galaxies, we also show the CCF of all galaxies of the compiled spec-$z$ catalog as a gray shade in Figure \ref{fig:IM_obscomp1}.
We find that the CCFs of the $\Mstar$--$10$, $\Mstar$--$9$, SFR--(ii), SFR--(iii), sSFR--(i), S16-sSFR--(ii), and SFG subsamples are similar to that of the all galaxies of the compiled spec-$z$ catalog. 
This is because that star-forming galaxies with $\Mstar \sim10^{10}$ M$_\sun$ and SFR $\sim 10-100$ M$_\sun$ yr$^{-1}$ are dominant in our compiled spec-$z$ catalog (see also Table \ref{tab:num_photoz}) and thus are responsible for the CCF in Figure \ref{fig:IM_specall}.
The other subsamples ($\Mstar$--$11$, $\Mstar$--$8$, SFR--(i), SFR--(iv), sSFR--(iii)) show a large difference between L16 and S16. We will discuss its reason in Sections \ref{sec:nature_all} and  \ref{sec:dis_absent_trend}.



\begin{figure*}
	\begin{center}
	\includegraphics[width=\linewidth]{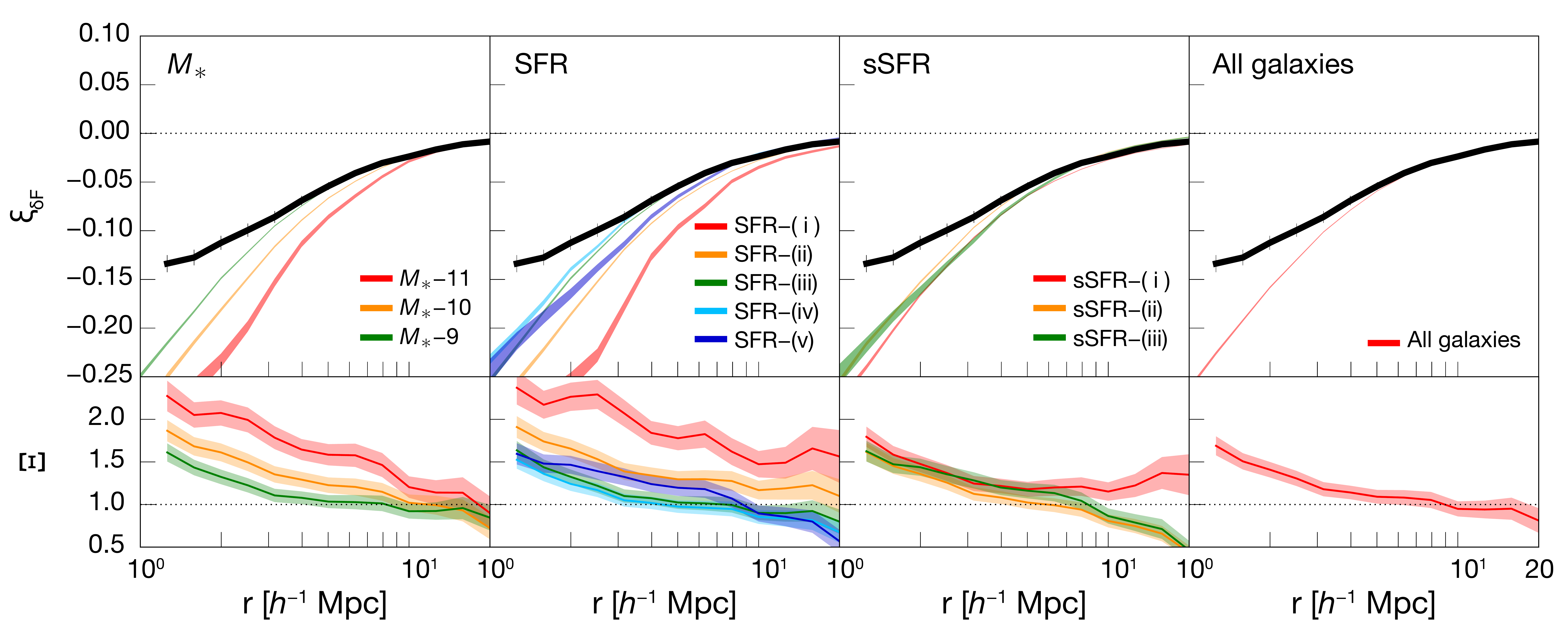} 
	\caption{
	($Top$) CCFs as a function of radius in comoving units from cosmological hydrodynamical simulations in \citet{momose20a} together with that of continuum-selected galaxies in Figure \ref{fig:IM_specall} which is colored in black.
    ($Bottom$) 
    $\Xi=\xi_\text{sim}$/$\xi_\text{all}$.
    }
	\label{fig:xi_sim}
	\end{center}
\end{figure*}


\subsubsection{Comparison of the CCF of Continuum-selected Galaxies with Those in Cosmological Hydrodynamical Simulations}
\label{sec:result_all_comp}

Because we do not find any significant $\Mstar$, SFR, and/or sSFR dependence in the CCFs, we compare the CCF of the continuum-selected galaxies to those obtained from cosmological hydrodynamical simulations
\citep{momose20a}.

In Figure \ref{fig:xi_sim}, we overlay the CCF of continuum-selected galaxies (black solid line) on those of $\Mstar$, SFR, and sSFR subsamples in \citet{momose20a} (colored lines).
The definition of each subsample and its name are the same as given in Table \ref{tab:num_photoz} (for more details, see Table 1 of \citealp{momose20a}).
We also show the CCF ratios,
$\Xi=\xi_\text{sim}/\xi_\text{all}$,
at the bottom of Figure \ref{fig:xi_sim}.

We find that the CCF of continuum-selected galaxies agrees with that of all simulated galaxies in \citet{momose20a} over $r=4-20$ $h^{-1}$ Mpc.
A detailed comparison with the $\Mstar$-, SFR-, and sSFR-dependent CCFs shows that the $\Mstar$--9, SFR--(iii), SFR--(iv), and sSFR--(ii) subsamples match well with the continuum-selected sample, particularly over $r=3-20$ $h^{-1}$ Mpc.
We briefly discuss this result in Section \ref{sec:nature_all}.



\begin{figure*}
	\begin{center}
	\includegraphics[width=\linewidth]{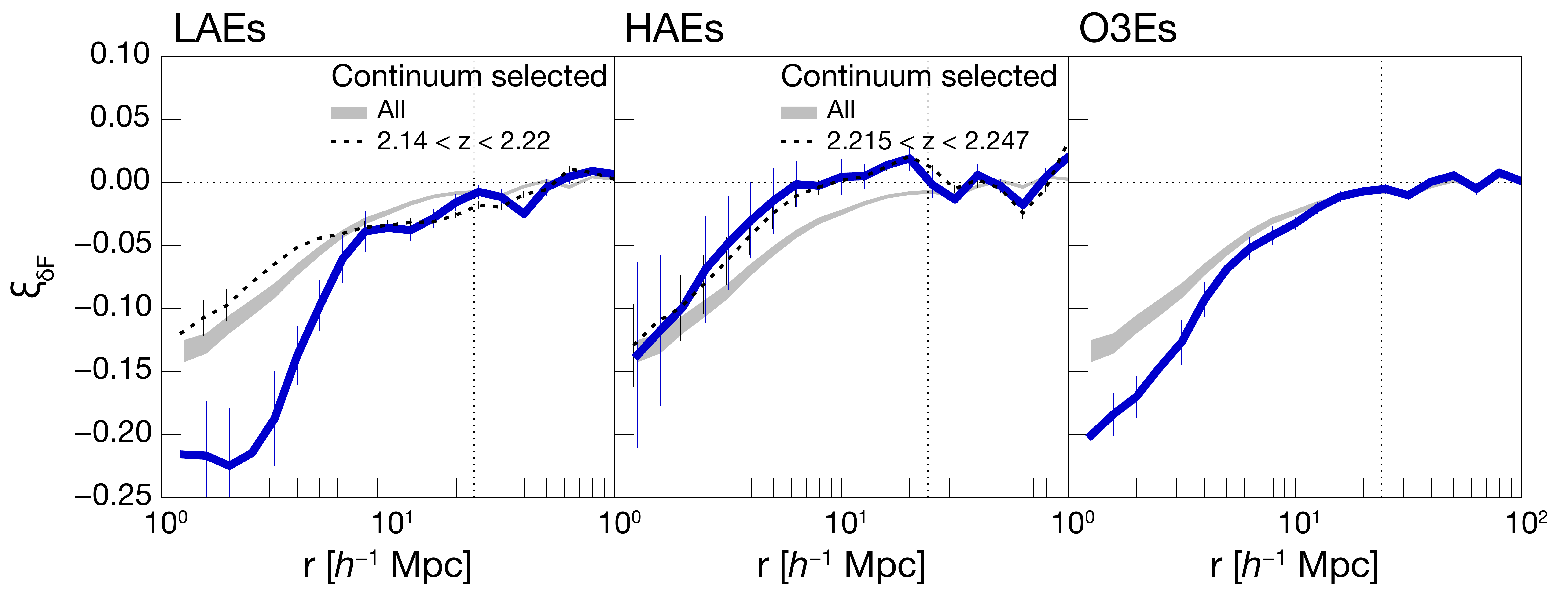} 
	\caption{
	CCFs of LAEs, HAEs, and O3Es from left to right as a function of radius in comoving units.
	Colored in gray is the CCF of the continuum-selected galaxies. Dashed lines in LAEs and HAEs indicate CCFs obtained from galaxies at the narrow redshift ranges defined by the FWHM of $NB$ filters.
	The meaning of vertical dotted lines is the same as in Figure \ref{fig:IM_specall}.}
	\label{fig:IM_galpop}
	\end{center}
\end{figure*}


\vspace{0.5cm}

\subsubsection{Galaxy Populations}
\label{sec:result_IM_galpop}

\paragraph{All Line Emitters}
Figure \ref{fig:IM_galpop} presents the CCFs of three different emitters of LAEs, HAEs, and O3Es from the left. We also plot the CCFs of the continuum-selected galaxies by a gray shade and the CCFs of continuum-selected galaxies within the redshift range defined by the FWHM of the $NB$ filter for LAEs at $2.14<z<2.22$ (HAEs at $2.215<z<2.247$) with a black dotted line. We refer to these redshift-specified CCFs of continuum-selected galaxies to compare the CCFs of LAEs and HAEs. Note that the CCFs of all continuum-selected galaxies from the entire CLAMATO redshift range and the above two specific redshift ranges have different slopes, although the $\xi_\text{$\delta$F}$ values at the center agree within the errors. We suspect that the differences in slope are due to cosmic variance \citep{momose20a}.

We detect signals for all the emitters. However, the strength and shape of the CCFs differ from each other. 
LAEs show the strongest signal among the three emitter populations with $\xi_\text{$\delta$F}=-0.23$, which is even stronger than the CCF of the continuum-selected galaxies at the same redshift range defined by the $NB$ filter \citep{naka12,konno16}. Additionally, LAEs' CCF is clearly different from those of any other galaxies, being flat up to $r\sim3$ $h^{-1}$ Mpc followed by a monotonic increase toward the cosmic mean. 
The CCF of O3Es in Figure \ref{fig:IM_galpop} (right) also shows the strongest signal with $\xi_\text{$\delta$F}=-0.21$ which is comparable to that of LAEs within the errors. The CCF monotonically increases up to the cosmic mean at $r\sim20$ $h^{-1}$ Mpc just like that of continuum-selected galaxies. On the other hand, the CCF of HAEs agrees well with that of continuum-selected galaxies at the same redshift range defined by the $NB$ filter \citep{sob13} in both the shape and the amplitude with $\xi_\text{$\delta$F}=-0.15$. This agreement indicates that HAEs are distributed in the cosmic web in the same manner as continuum-selected galaxies. 
We will discuss it in Sections \ref{sec:dis_HAE} and \ref{sec:dis_comppower}.


\begin{figure*}
	\begin{center}
	\includegraphics[width=\linewidth]{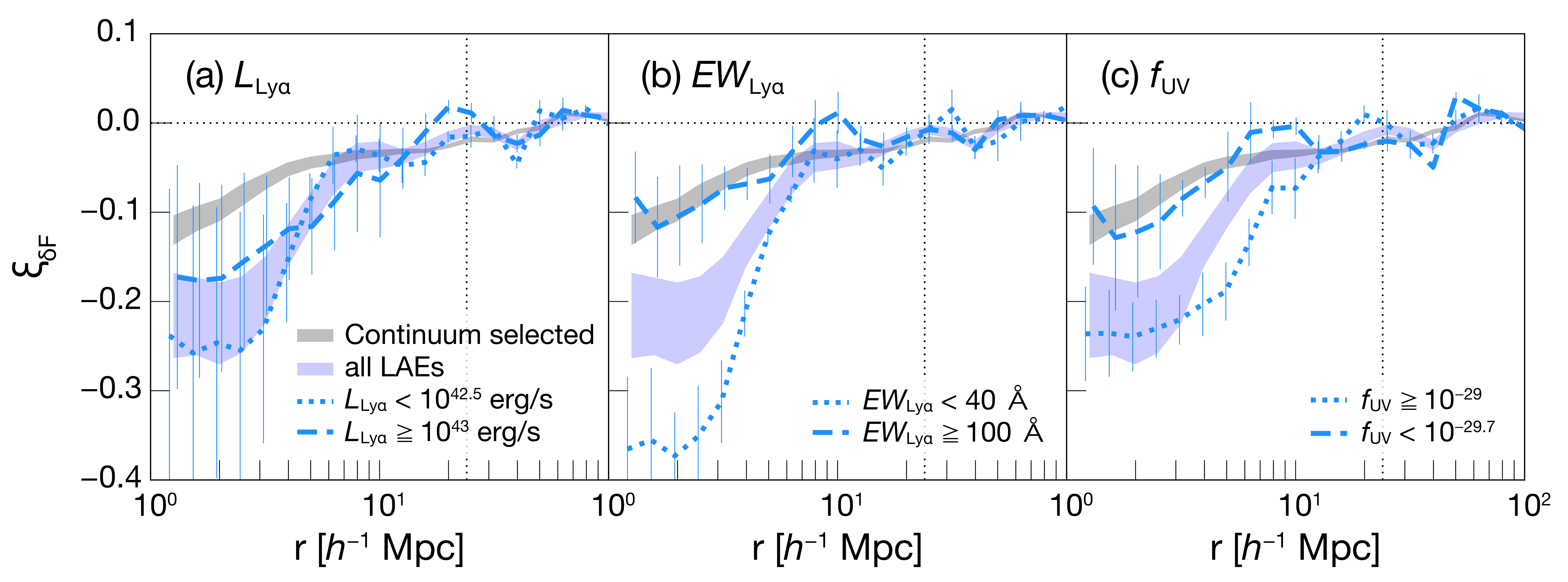} 
	\caption{
    CCFs of LAE subsamples divided by (a) $L_\text{Ly$\alpha$}$, (b) $EW_\text{Ly$\alpha$}$, and (c) UV luminosity as a function of radius in comoving units.
    The meaning of vertical dotted lines is the same as in Figure \ref{fig:IM_specall}.
    }
	\label{fig:IM_LAE}
	\end{center}
\end{figure*}


\paragraph{LAEs}
As we note in Section \ref{sec:catalogdata}, our LAEs are likely biased toward higher $L_\text{Ly$\alpha$}$, and thus their CCF in Figure \ref{fig:IM_galpop} (left) could more reflect the gas environments of such luminous LAEs. 
To evaluate the effect of this bias, we make two subsamples according to Ly$\alpha$ luminosity ($L_\text{Ly$\alpha$}$), 
an Ly$\alpha$-luminous one with $L_\text{Ly$\alpha$}\geq10^{43}$ erg s$^{-1}$ (four objects) and 
an Ly$\alpha$-faint one with $L_\text{Ly$\alpha$}<3\times10^{42}$ erg s$^{-1}$ (five objects), 
and calculate the CCF for each.
As found from Figure \ref{fig:IM_LAE} (a), both the Ly$\alpha$-luminous and Ly$\alpha$-faint subsamples have stronger signals than the continuum-selected galaxies. 
Although consistent within the errors, the Ly$\alpha$-faint subsample has a slightly stronger signal than the Ly$\alpha$-luminous one, with a flatter CCF up to $r\sim3$ $h^{-1}$ Mpc. 
These results perhaps indicate that LAEs with different $L_\text{Ly$\alpha$}$ are distributed in the cosmic web in a different manner.

To understand the IGM environments of LAEs in more detail, we conduct an additional investigation by making another four subsamples based on equivalent width ($EW_\text{Ly$\alpha$}$) and UV luminosity ($f_\text{UV}$): large-$EW_\text{Ly$\alpha$}$ ($EW_\text{Ly$\alpha$}\geq100$ {\AA}, seven objects), small-$EW_\text{Ly$\alpha$}$ ($EW_\text{Ly$\alpha$}<40$ {\AA}, three objects), UV-luminous ($f_\text{UV}\geq1\times10^{-29}$ erg cm$^{-2}$ s$^{-1}$ Hz$^{-1}$, four objects), and UV-faint ($f_\text{UV}<2\times10^{-30}$ erg cm$^{-2}$ s$^{-1}$ Hz$^{-1}$, four objects). The CCFs of these four subsamples are shown in Figure \ref{fig:IM_LAE} (b) and (c).
For the \textit{EW} subsamples, we find a clear difference in their CCFs. The CCF of the small-$EW_\text{Ly$\alpha$}$ subsample has a similar shape to that of all LAEs, but with a stronger signal of $\xi_\text{$\delta$F}=-0.37$, while the CCF of the large-$EW_\text{Ly$\alpha$}$ subsample shows a good agreement with that of continuum-selected galaxies. 
For the UV subsamples, the UV-luminous one has a stronger signal than the UV-faint one and shows a flat CCF up to as large as $r\sim5$ $h^{-1}$ Mpc. 
On the other hand, the UV-faint subsample has a similar CCF shape and strength to the continuum-selected galaxies. 



\begin{figure*}
	\begin{center}
	\includegraphics[width=0.8\linewidth]{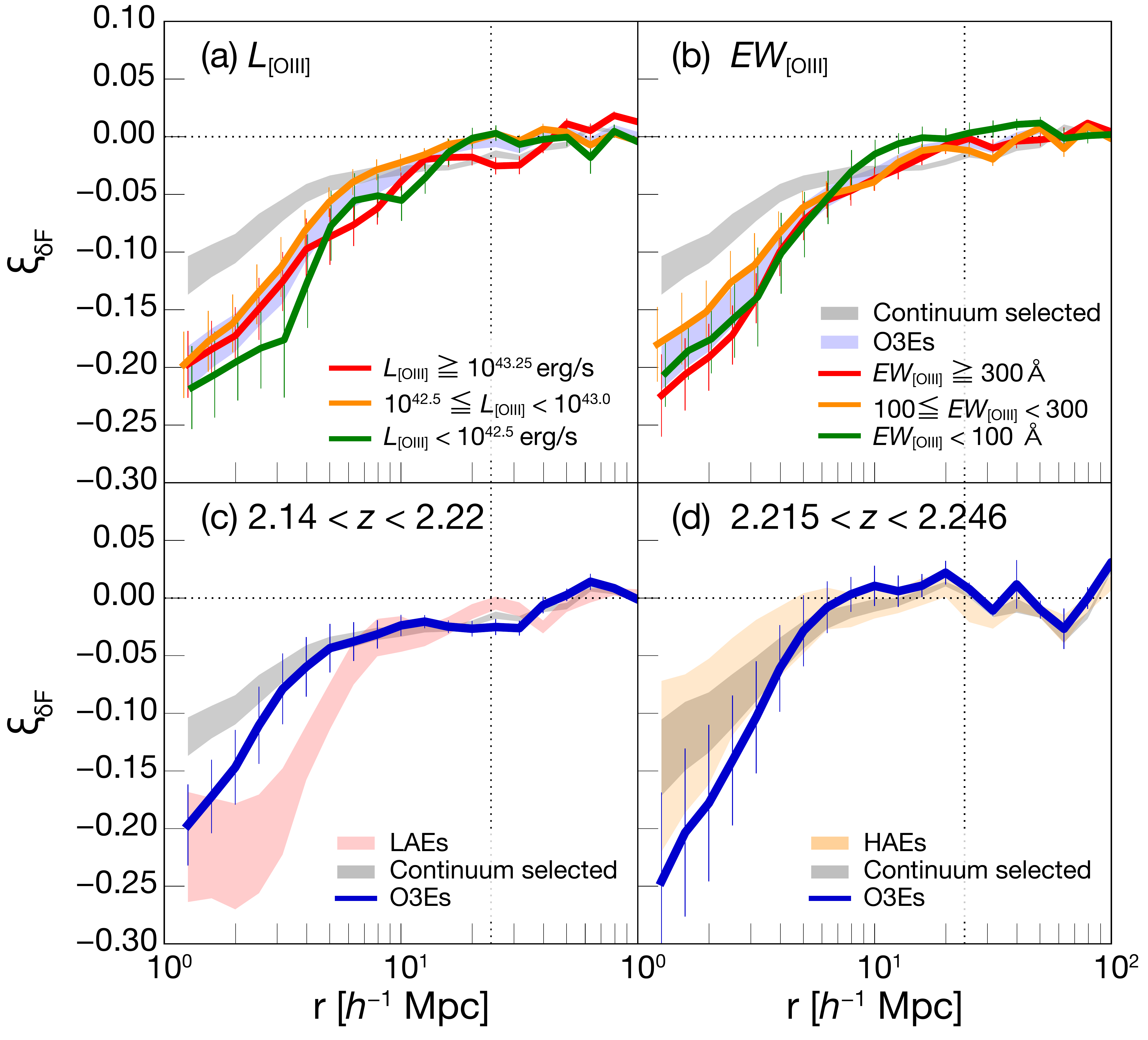} 
	\caption{
    CCFs between O3Es and Ly$\alpha$ absorption as a function of radius in comoving units.
    {\it Panel (a):} $L_\text{[\sc Oiii]}$ subsamples. Gray and royal blue shades represent the CCFs of continuum-selected galaxies in the compiled spec-$z$ catalog and all O3Es, respectively.
    {\it Panel (b):} $EW_\text{[{\sc Oiii}]}$ subsamples. Meanings of gray and royal blue shades are the same as in Panel (a).
    {\it Panel (c):} The O3E subsample is limited to the same redshift range as of LAEs, $2.14<z<2.22$. The CCF of LAEs is colored in pink. A gray shade represents the CCF of continuum-selected galaxies at $2.14<z<2.22$.
    {\it Panel (d):}
    For the O3E subsample in the same redshift range as of HAEs, $2.215<z<2.246$. The CCF of HAEs is colored in orange. A gray shade represents the CCF of continuum-selected galaxies at $2.215<z<2.246$.
    The meaning of vertical dotted lines is the same as in Figure \ref{fig:IM_specall}.
    }
	\label{fig:IM_O3E}
	\end{center}
\end{figure*}


\paragraph{O3Es}
Similar to the LAEs, our O3Es are also biased toward higher $L_\text{[\sc Oiii]$\lambda\lambda5007$}$. 
We measure CCFs by dividing our O3Es into three subsamples based on $L_\text{[\sc Oiii]$\lambda\lambda5007$}$ and $EW_\text{[\sc Oiii]$\lambda\lambda5007$}$ in Figure \ref{fig:IM_O3E} (a) and (b). 
We do not find any clear dependence on either $L_\text{[\sc Oiii]$\lambda\lambda5007$}$ or $EW_\text{[\sc Oiii]$\lambda\lambda5007$}$.

We also compare the CCF of O3Es with those of LAEs, HAEs, and continuum-selected galaxies at $2.14<z<2.22$ and $2.215<z<2.247$. The difference from Figure \ref{fig:IM_galpop} (right) is that only O3Es in those redshift ranges are used.
As shown in Figure \ref{fig:IM_O3E} (c) and (d), the O3Es have a stronger CCF than the HAEs and a slightly weaker CCF than the LAEs.

\paragraph{AGNs and SMGs}


\begin{figure*}
	\begin{center}
	\includegraphics[width=\linewidth]{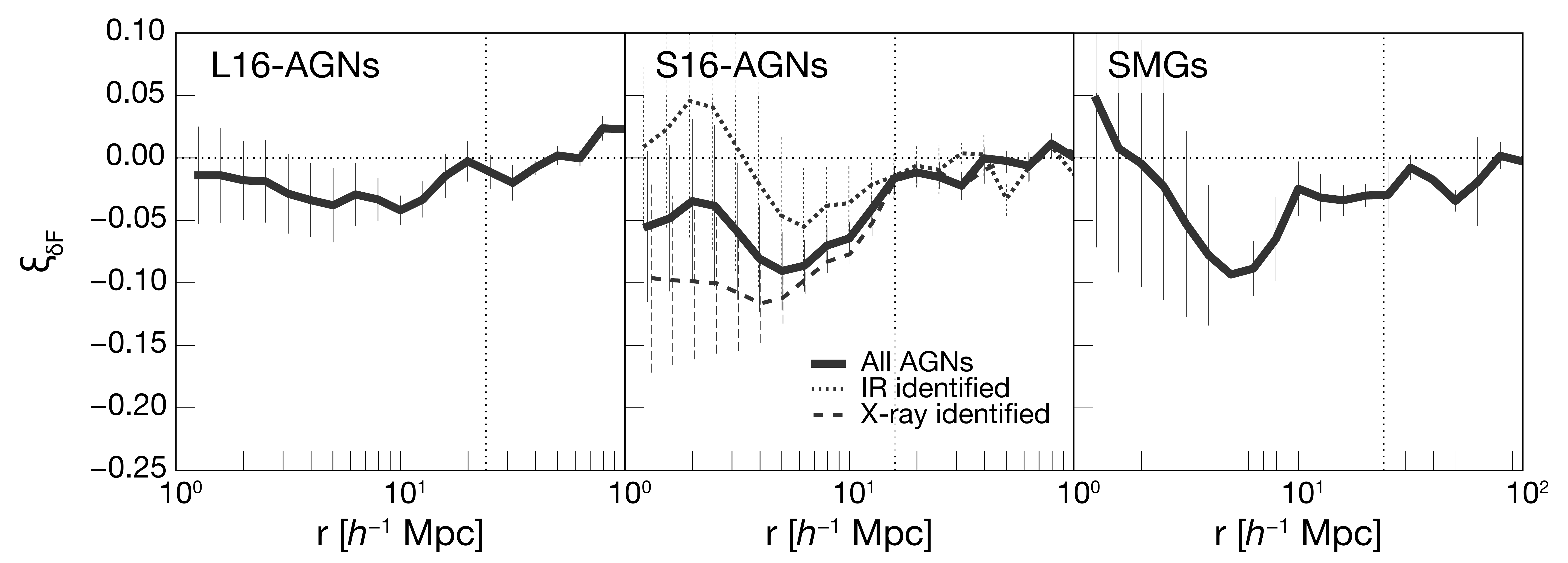} 
	\caption{
	CCFs of L16-AGNs, S16-AGNs, and SMGs from left to right as a function of radius in comoving units.
    The meaning of vertical dotted lines is the same as in Figure \ref{fig:IM_specall}.
    }
	\label{fig:IM_agns}
	\end{center}
\end{figure*}


The CCFs of AGNs and SMGs are shown in Figure \ref{fig:IM_agns}, which are greatly different from those of continuum-selected galaxies and emitters. 
A common feature of the CCFs of the L16-AGNs and the S16-AGNs is a negative peak (i.e., the largest signal) at $r\sim5$ $h^{-1}$ Mpc. Although this feature of the L16-AGNs is weak, being only $<2\sigma$ away from the cosmic mean ($\xi_\text{$\delta$F}=0$), that of the S16-AGNs is significant with $\xi_\text{$\delta$F}=-0.1$. 
The CCF of SMGs also shows a negative peak at $r\sim5-6$ $h^{-1}$ Mpc with $\xi_\text{$\delta$F}=-0.09$. 
Because two of the four SMGs have an X-ray source within a $1\arcsec$ aperture, and one of the remaining two has an X-ray source within a $2\arcsec$ aperture, most of our SMG sample are AGNs. Thus, the strongest CCF signal at $r\sim5-6$ $h^{-1}$ Mpc away from the center seen in both AGNs and SMGs is probably due to the AGN activity of the central black hole and thus may be a general feature of the IGM {\sc Hi} around AGNs. 
\citet{mukae19} have also found $5-10$ $h^{-1}$ Mpc off-center negative peaks around QSOs \citep{mukae19}. 
Further discussion is given in Section \ref{sec:dis_AGNSMG}.

Figure \ref{fig:IM_agns} also shows an interesting trend depending on the AGN type. $X$-ray-identified AGNs, which are all L16-AGNs and a fraction of the S16-AGNs (dashed line), show a negative value at the center and a decrease up to $r\sim4-5$ $h^{-1}$ Mpc until reaching the negative peak. 
On the other hand, S16's IR-identified AGN and SMGs have a positive $\xi_\text{$\delta$F}$ value at the center.   
It implies that $X$-ray-identified AGNs have a slightly stronger CCF than IR-identified one. 
We also discuss it in Section \ref{sec:dis_AGNSMG}.



\begin{figure*}
	\begin{center}
	\includegraphics[width=\linewidth]{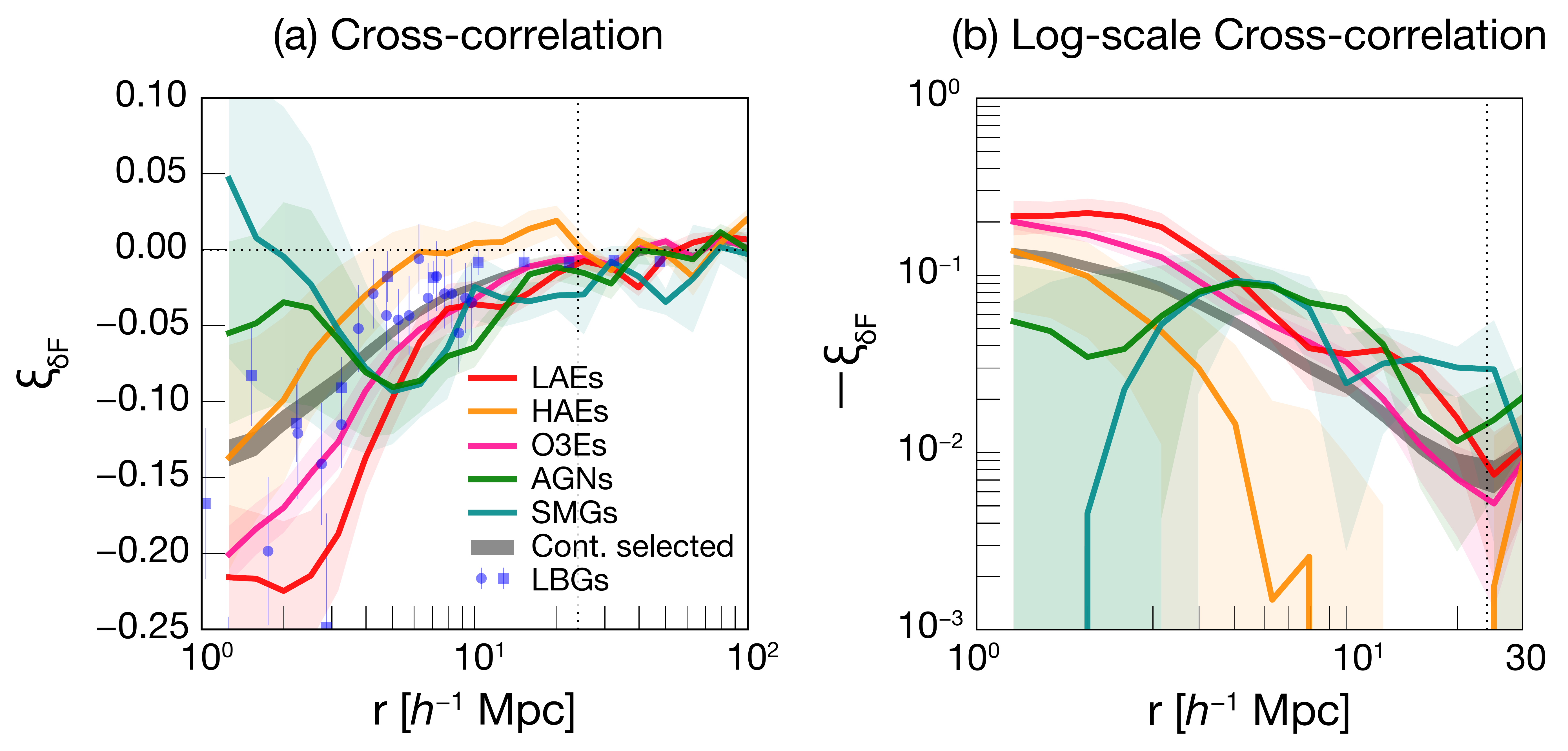} 
	\caption{
    CCFs of different galaxy populations plotted on linear scale (a) and log scale (b) as a function of radius in comoving units.
    The CCF between Ly$\alpha$ absorption and LBGs from the literature is also plotted by blue circles \citep{adelberger05} and squares \citep{bielby17}.
    The meaning of vertical dotted lines is the same as in Figure \ref{fig:IM_specall}.
    }
	\label{fig:IM_all_galpop}
	\end{center}
\end{figure*}


\subsubsection{Comparison of the CCFs}
\label{sec:res_IMcomp}

We show the CCFs of all galaxy populations and all continuum-selected galaxies simultaneously in Figure \ref{fig:IM_all_galpop} (a) together with Ly$\alpha$ forest--LBGs CCFs evaluated at $z\sim3$ in the literature \citep{adelberger05,bielby17}. 
For visibility purposes, we only plot the CCF of the S16-AGNs as the representative of AGNs.

The CCF of continuum-selected galaxies agrees well with that of LBGs at $r>3$ $h^{-1}$ Mpc, though the latter is largely scattered within $r<3$ $h^{-1}$ Mpc. 
However, the CCFs of other galaxy populations show a variation. In addition, all galaxy populations except HAEs have a stronger signal than the continuum-selected galaxies at $r>5$ $h^{-1}$ Mpc in Figure \ref{fig:IM_all_galpop} (b).

To quantify the CCFs, we fit them by a power-law with: 
\begin{equation}
    \xi_\text{$\delta$F}(r) = \left( \frac{r}{r_0} \right)^\gamma,
\end{equation} 
where $r_0$ and $\gamma$ are the clustering length and slope, respectively. We fit the CCFs of star-forming galaxies (i.e., continuum-selected galaxies, LAEs, HAEs, O3Es) over $r=3-24$ $h^{-1}$ Mpc; $3$ $h^{-1}$ Mpc corresponds to the spectral resolution of the CLAMATO and $24$ $h^{-1}$ Mpc corresponds to the CLAMATO's short side on the sky.
For AGNs and SMGs, we fit their CCFs over $r=5-24$ $h^{-1}$ Mpc, because the observed CCFs deviate from a single power-law over $3<r<5$ $h^{-1}$ Mpc. 
The best-fit parameters are summarized in Table \ref{tab:IM_bestfit}. 

We find the best-fit parameters of continuum-selected galaxies to be $r_0=0.45\pm0.04$ and $\gamma=1.23\pm0.04$, which are comparable to those evaluated in the literature.
\citet{bielby17} have calculated a CCF between Ly$\alpha$ absorption and LBGs at $z\sim3$ and obtained its best-fit parameters to be $r_0=0.27\pm0.14$ and $\gamma=1.1\pm0.2$. \citet{tejos14} have examined the variety of the CCFs for galaxies at $z<1$ depending on the {\sc Hi} column density of the IGM and found the best-fit parameters of weak {\sc Hi} systems ($N_\text{{\sc Hi}}<10^{14}$ cm$^{-2}$) to be $r_0=0.2\pm0.4$ and $\gamma=1.1\pm0.3$.
However, \citet{momose20a} have obtained slightly larger values of $r_0=0.62\pm0.04$ and $\gamma=1.37\pm0.04$ over the same fitting range\footnote{
\citet{momose20a} have evaluated the best-fit parameters ($r_0$, $\gamma$) of the IGM regime over $r=0-10$ $h^{-1}$ Mpc and obtained ($0.34\pm0.03$, $1.07\pm0.04$). However, we recalculate them over $r=3-24$ $h^{-1}$ for a fair comparison with those of this study.}.
We suspect that the slightly smaller best-fit parameters in this study may be due to the smearing of the CCF because of the lower effective spectral resolution of the CLAMATO than that of the simulations \citep{momose20a}, whose line-of-sight resolution is $0.4$ $h^{-1}$ Mpc at $\langle z \rangle=2.3$.

All galaxy populations except HAEs are found to have a similar slope to that of continuum-selected galaxies with $\gamma=1.2-1.3$. Nonetheless, their clustering lengths $r_0$ are larger than those of continuum-selected galaxies. 
On the other hand, the best-fit parameters of HAEs agree within the errors with those of continuum-selected galaxies in the same redshift range.


\begin{table*}
    \caption{The best fit parameters of our CCFs.}
\begin{center}
\begin{tabular}{lrcc}
\tableline
    Sample & $r_0$ [$h^{-1}$ Mpc] & $\gamma$ & Fitting Range [$h^{-1}$ Mpc]  \\
\hline
    Continuum-selected galaxies & $0.46\pm0.04$ & $1.23\pm0.04$ & $3-24$   \\
\hline
    LAEs         & $0.78\pm0.13$ & $1.25\pm0.10$ & $3-24$   \\
    Galaxies at $2.14<z<2.22$ & ($2.92\pm2.40$)$\times10^{-3}$ & $0.41\pm0.04$ & $3-24$ \\
\hline
    HAEs         & $1.24\pm0.23$ & $3.16\pm0.55$ & $3-24$   \\
    Galaxies at $2.215<z<2.247$ & $1.11\pm0.17$ & $2.57\pm0.31$ & $3-24$    \\
\hline
    O3Es         & $0.72\pm0.09$ & $1.37\pm0.08$ & $3-24$   \\
\hline
    S16-AGN      & $0.99\pm0.42$ & $1.32\pm0.26$ & $5-24$   \\
    SMGs         & $0.41\pm0.26$ & $0.94\pm0.19$ & $5-24$   \\
\hline
\end{tabular}
\label{tab:IM_bestfit}
\end{center}
\end{table*}



\begin{figure*}
	\begin{center}
	\includegraphics[width=\linewidth]{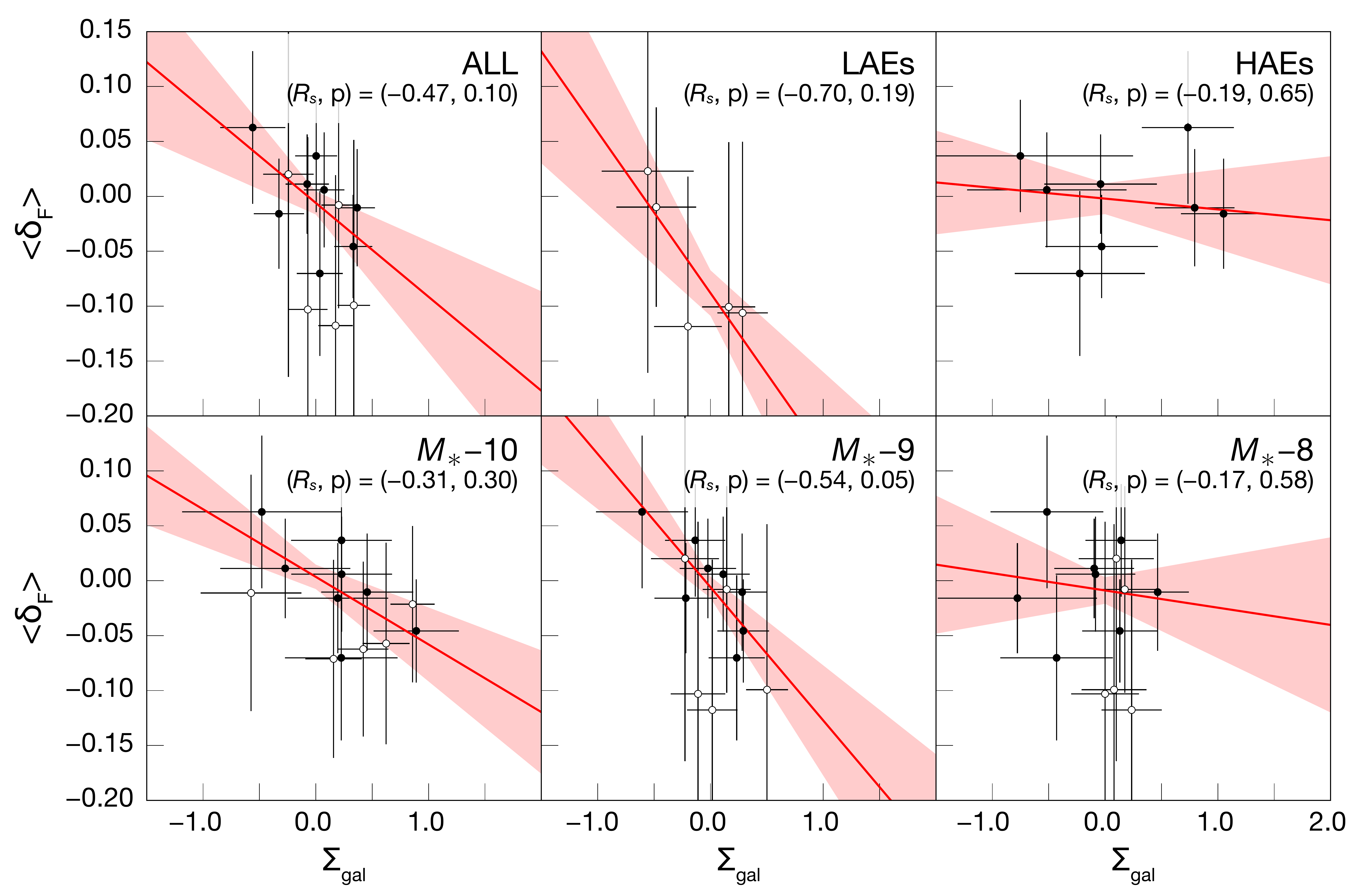} 
	\caption{
    $\langle \delta_\text{F} \rangle - \Sigma_\text{gal}$ relations obtained from local minima and maxima. 
    Points colored in white and black are from $2.14<z<2.22$ and $2.215<z<2.247$ ranges, respectively.
    The best-fit linear models obtained from the combined two redshift ranges are shown by thick red lines with errors.
    } 
	\label{fig:den_bias}
	\end{center}
\end{figure*}

\begin{figure*}
	\begin{center}
	\includegraphics[width=\linewidth]{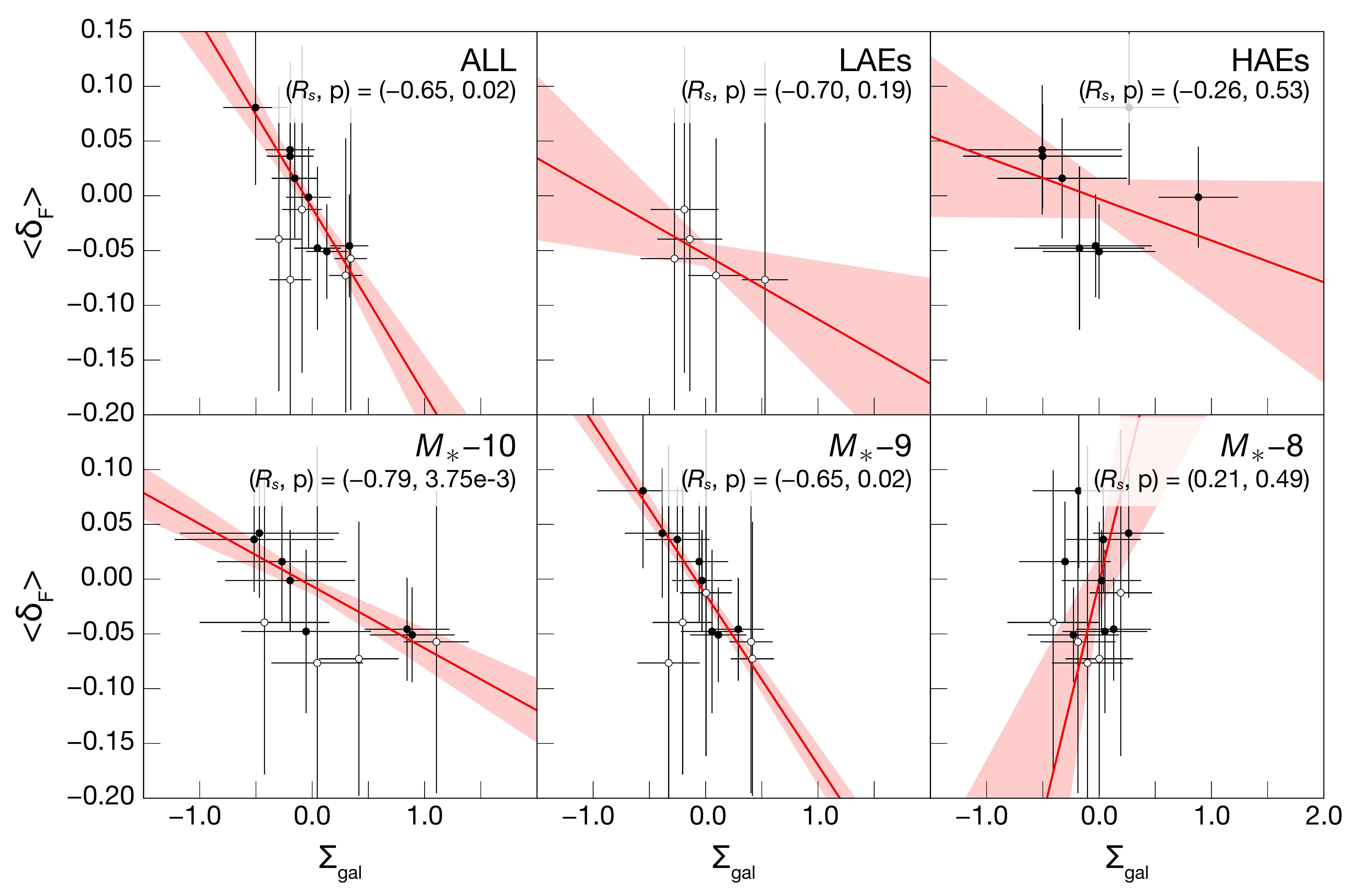} 
	\caption{
    Same as Figure \ref{fig:den_bias}, but evaluated from random positions.}
	\label{fig:den_unbias}
	\end{center}
\end{figure*}


\subsection{Overdensity Analysis}
\label{sec:result_mukae}

Overdensity analysis is conducted on photo-$z$ galaxies in L16 (ALL) and $NB$-selected LAEs and HAEs. We divide the photo-$z$ galaxies into three subsamples based on stellar mass 
($\Mstar \geq10^{10}$ M$_\sun$: L16-$\Mstar$--$10$, $10^9\leq \Mstar$/M$_\sun <10^{10}$: L16-$\Mstar$--$9$, $10^8\leq \Mstar$/M$_\sun < 10^9$: L16-$\Mstar$--$8$). The number of galaxies in individual subsamples is shown in Tables \ref{tab:overdensity} or \ref{tab:overdensity_ran}.  
Figure \ref{fig:den_bias} presents $\langle \delta_\text{F} \rangle$--$\Sigma_\text{gal}$ relations for the six subsamples. We assume Poisson noise to evaluate the error in $\Sigma_\text{gal}$ measurements. 

We first find a negative $\langle \delta_\text{F} \rangle$--$\Sigma_\text{gal}$ correlation for all six subsamples by eyes. 
To assess the significance of the correlation, we calculate the Spearman's rank correlation coefficient $R_s$ for each subsample, as summarized in Table \ref{tab:overdensity}. We find negative $R_s$ values for all the subsamples with $R_s=-0.17$ to $-0.70$. However, considering larger $p-$values ($0.19-0.65$) except for the L16-$\Mstar$--$9$ and ALL, all the anti-correlations in Figure \ref{fig:den_bias} are statistically insignificant. 
For the L16-$\Mstar$--$9$ and ALL, we obtain Spearman's coefficients of $R_s=-0.54$ with the $95\%$ confidence level and $R_s=-0.47$ with the $90\%$ confidence level, indicating the presence of weak anti-correlations. 
Likewise, \citet{mukae17} have also found an anti-correlation in their $\langle \delta_\text{F} \rangle$--$\delta_\text{gal}$ relation evaluated from a cylinder that has a radius of $r=8.8$ $h^{-1}$ Mpc and the length of $88$ $h^{-1}$ Mpc at $z=2.5$, at a $\sim90\%$ confidence level with $R_s=-0.39$. 
The same analysis performed for all simulated galaxies in \citet{momose20a} gives $R_s=-0.37$ with the $98\%$ confidence level, indicating a weak anti-correlation in the $\langle \delta_\text{F} \rangle$--$ \Sigma_\text{gal}$ relation.

We also apply $\chi$-square fitting to the $\langle \delta_\text{F} \rangle$--$\Sigma_\text{gal}$ relations of all of our subsamples shown in Figure \ref{fig:den_bias} with a linear model of
\begin{equation}
  \langle \delta_\text{F} \rangle = \alpha + \beta\,  \Sigma_\text{gal}.
\end{equation}
The best-fit parameters are summarized in Table \ref{tab:overdensity}.
Those of the two subsamples with a statistically confirmed anti-correlation are ($\alpha=-0.006\pm0.010$, $\beta=-0.121\pm0.040$) for L16-$\Mstar$--$9$ and ($\alpha=-0.006\pm0.011$, $\beta=-0.085\pm0.040$) for ALL.
\citet{mukae17} have obtained $\alpha=-0.17\pm0.06$ and $\beta=-0.14^{+0.06}_{-0.16}$. 
\citet{momose20a} have also evaluated the best-fit linear model of $\langle \delta_\text{F} \rangle$--$\Sigma_\text{gal}$ relations with the numerical simulations and found $\alpha=-0.126\pm0.006$ and $\beta=-0.018\pm0.006$. 
The slopes $\beta$ of L16-$\Mstar$--$9$ and ALL are consistent with the one by \citet{mukae17} within the errors but are much steeper than the one obtained from the simulations.
As has already been discussed in \citet{momose20a}, such a discrepancy in slope between observations and simulations may be due to photo-$z$ errors. 
For example, the typical photo-$z$ error of L16 galaxies ($\sigma_z = 0.07$) is larger than the thickness of the HAE slice ($\Delta z=0.032$), meaning that $\Sigma_\text{gal}$ measurements have been contaminated from galaxies outside the slice. This smearing of $\Sigma_\text{gal}$ would make an observed slope steeper than the true value.

Since \citet{mukae17} have pointed out a possible bias in the $\langle \delta_\text{F} \rangle$--$\Sigma_\text{gal}$ relation due to the position of sightlines, we also perform overdensity analysis based on randomly selected sightline positions.
As shown in Figure \ref{fig:den_unbias} and Table \ref{tab:overdensity_ran}, similar results are obtained for most of the subsamples.
Statistically significant anti-correlations are confirmed only in L16-$\Mstar$--$10$, L16-$\Mstar$--$9$, and ALL with $98\%$, more than $99\%$, and $98\%$ confidence levels respectively, which are also consistent with the literature \citep{mukae17,momose20a}.

Separately from the significance of correlations, we also find intriguing results for LAEs. Although the error bars are large, the distribution of LAEs in $\Sigma_\text{gal}$ seems to be slightly wider than those of L16-$\Mstar$--$9$ and L16-$\Mstar$--$8$ subsamples at $2.14<z<2.22$ (white points in Figures \ref{fig:den_bias} and \ref{fig:den_unbias}). The width of the $\Sigma_\text{gal}$ distribution is $0.84$, $0.73$, and $0.47$ ($0.80$, $0.74$, and $0.41$) for LAEs, L16-$\Mstar$--$9$, and L16-$\Mstar$--$8$ subsamples in Figure \ref{fig:den_bias} (Figure \ref{fig:den_unbias}). We will briefly discuss possible implications from the larger $\Sigma_\text{gal}$ distribution in Section \ref{sec:dis_LAEHAE}.


\section{Discussion}


\subsection{What Can We Find About the IGM--Galaxy Connection through the Two Approaches?}
\label{sec:dis_method}

In this study, we adopt two methods to investigate the IGM--galaxy connection: one is cross-correlation analysis, and the other is overdensity analysis. These methods are sensitive to different aspects of the IGM--galaxy connection and have both strong and weak points. 


\subsubsection{Cross-correlation Analysis}
\label{sec:dis_IMmethod}

By measuring average {\sc Hi} overdensities as a function of distance, cross-correlation analysis tells us how a given galaxy population traces the cosmic {\sc Hi} web. The advantage of this method is that a CCF signal can be detected even for a small number of galaxies. Indeed, as shown in Section \ref{sec:result_IM}, we confirm a significant CCF signal for only seven HAEs and four SMGs. 

Note, however, that a CCF from a small sample may be greatly different from the true one owing to large statistical errors \citep{momose20a}.
The irregular shapes seen in the CCFs of $\Mstar$--$11$, $\Mstar$--$8$, SFR--(i), SFR--(iv), and sSFR--(iii) subsamples could be due to their small sample sizes.
In contrast, the CCF obtained from a sufficiently large number of randomly selected galaxies is close to the true one.
It can be the case for the $\Mstar$--$9$, $\Mstar$--$10$, SFR--(ii), sSFR--(i), and SFG subsamples, which show consistent CCFs between L16 and S16.
In addition, we can obtain a CCF similar to the true one even from a small sample, if galaxies of a given type reside in a similar gas environment. 
It may be the case for LAEs, HAEs, AGNs, and SMGs.

A disadvantage of cross-correlation analysis is that it requires spec-$z$ measurements.
This is because the typical photo-$z$ error (i.e., $\sigma_z=0.05-0.1$ at $z\sim2$ corresponding to $40-90$ $h^{-1}$ Mpc, e.g., \citealp{muzzin13,laigle16,straa16}) is much larger than the scales over which the cosmic {\sc Hi} density varies (a few Mpc scale). If galaxies with spec-$z$ measurements do not represent the parent sample, the CCF obtained from them may be biased in some manner.


\subsubsection{Overdensity Analysis}

The other method used in this study is overdensity analysis.
An advantage of this method is that it can evaluate the tightness of the correlation between galaxy and IGM densities (the IGM--galaxy connection) for a given size of cells. If sufficiently long (along the line-of-sight) cells are adopted as in the case of this study ($\Delta z=0.08$ and $0.032$), photo-$z$ samples can be used. A drawback of using such long cells is that the overdensities of galaxies and IGM for such cells are small and hence noisy. Owing to this disadvantage, combined with the fact that the sky coverage of the CLAMATO is not large enough to put many independent cells, we cannot confirm a $\langle \delta_\text{F} \rangle$--$\Sigma_\text{gal}$ correlation with a high significance for several subsamples.


\begin{figure*}
	\begin{center}
	\includegraphics[width=0.75\linewidth]{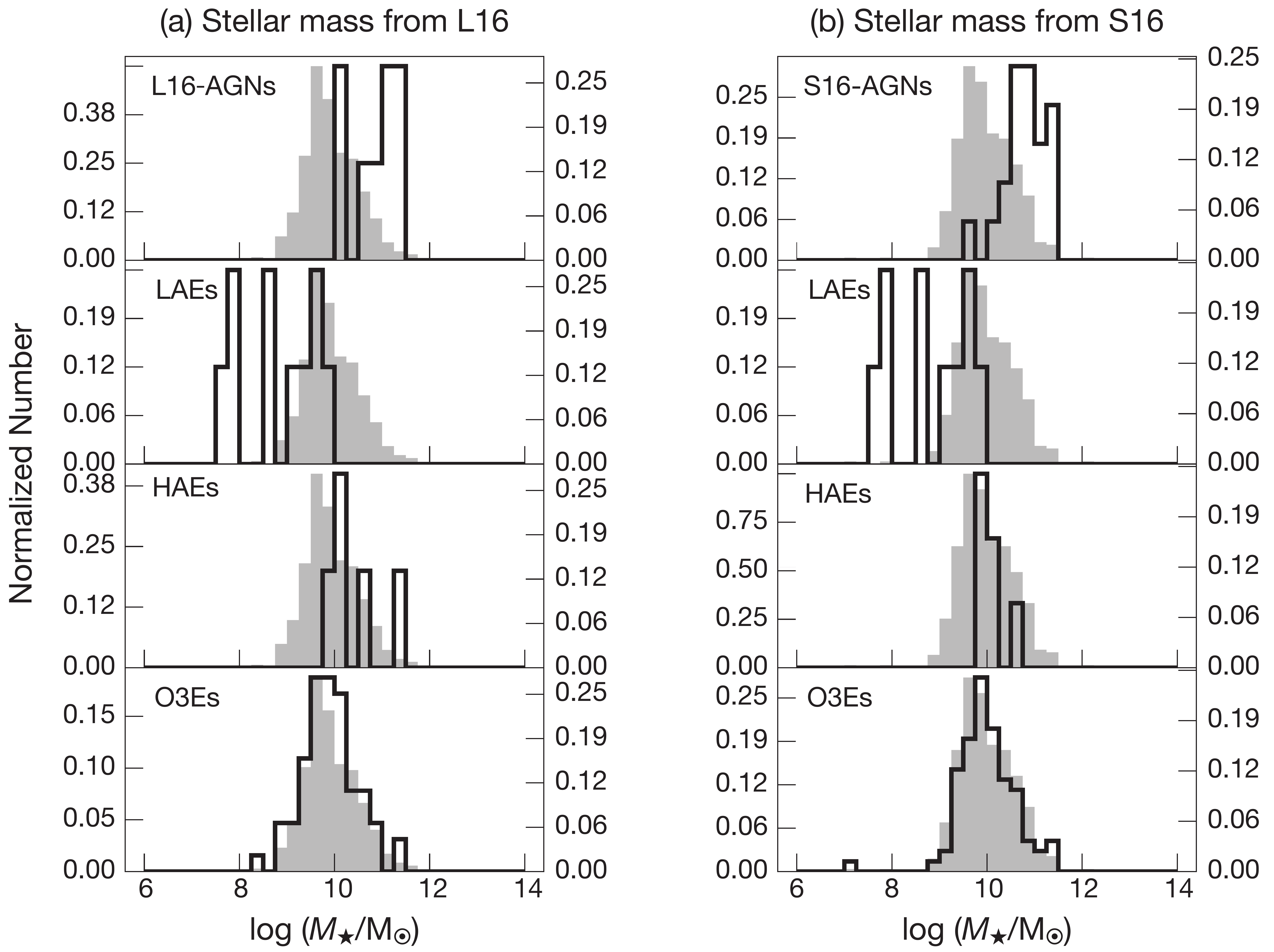} 
	\caption{
    Normalized number histograms of galaxies as a function of stellar mass ($M_\star$). The solid black line indicates the histogram of each galaxy population -- AGNs, LAEs, HAEs, and O3Es from top to bottom. The histograms of the L16 and S16 spec-$z$ samples are overlaid by gray shades in panels (a) and (b) for comparison. The normalized numbers of each galaxy population and L16/S16 are shown on the left and right axes, respectively. Note that we only plot galaxies with $\Mstar$ estimates.
    } 
	\label{fig:hist_Ms}
	\end{center}
\end{figure*}


\subsection{Implications for the IGM--Galaxy Connection of Each Galaxy Population}
\label{sec:IGM--galaxy}

In this subsection, we discuss the IGM--galaxy connection depending on galaxy properties (i.e., $\Mstar$, SFR, and sSFR) and galaxy populations.



\subsubsection{Nature of the Major Contributor to the CCF of Continuum-selected Galaxies}
\label{sec:nature_all}

From Section \ref{sec:result_IM_galpro}, we find that galaxies with  $\Mstar\sim10^{10}$ M$_\odot$ and SFR $\sim10$ M$_\odot$ yr$^{-1}$ are dominant and responsible for the CCF of continuum-selected galaxies. 
In fact, the number histograms of $\Mstar$ for the L16 and S16 samples shown by gray shades in Figure~\ref{fig:hist_Ms} have a peak at around $\Mstar\sim10^{10}$ M$_\odot$.
On the other hand, the comparison with \citet{momose20a} in Figure \ref{fig:xi_sim} shows that the CCF of continuum-selected galaxies is reproduced by the $\Mstar$--9, SFR--(iii), SFR--(iv), and sSFR--(ii) subsamples, implying that continuum-selected galaxies have $\Mstar\sim10^9$ M$_\odot$ and SFR $\sim0.1-1$ M$_\odot$ yr$^{-1}$ (see also Section \ref{sec:result_all_comp}). These small discrepancies in $\Mstar$ and SFR between the observed and simulated galaxies, if real, may be due to differences in galaxy models used in SED fitting between the L16/S16 samples and the simulations.
Unfortunately, however, we cannot identify the cause at this point, because the volume covered by the CLAMATO is still insufficient. 
Future surveys for 3D tomography, such as the one by the Prime Focus Spectrograph (PFS) on the Subaru Telescope, will enable us to investigate it in detail.

\subsubsection{Reasons for the Lack of CCF Variation in $\Mstar$, SFR, and sSFR Subsamples}
\label{sec:dis_absent_trend}

Although \citet{momose20a} have found a clear dependence of the CCF on $M_\star$ and SFR, we do not find such significant dependence in this study (see Section \ref{sec:result_IM_galpro}).
We also do not find any clear dependence on galactic properties, even though we divide the galaxy sample into two (see Appendix~\ref{app:IM_div2}). 
We give two possible reasons in the following. 

First is the small sample size used in our observational analysis. 
\citet{momose20a} have demonstrated that a small randomly selected sample cannot always reproduce the true CCF.
As we have already described in Section \ref{sec:dis_method}, such a small sample size can cause an irregular CCF like those of several subsamples in Figure \ref{fig:IM_obscomp1}.

The second possible reason is the errors in stellar mass and SFR estimates for the L16 and S16 samples. Fainter galaxies have more chance to be assigned to wrong subsamples owing to larger photometric errors. Furthermore, L16 and S16 use different SED models, implying that galaxies with the same SED measurements can even be assigned to different subsamples. Smaller subsamples will suffer more from such misclassification because of heavier contaminations from other subsamples. Indeed, the subsamples that give consistent CCFs between L16 and S16 have relatively large sizes.
In order to examine whether or not the mass and SFR dependence of the CCF found in \citet{momose20a} exist in the real observational data, we need a larger galaxy sample and/or a larger 3D tomography volume.

\subsubsection{LAEs}
\label{sec:dis_LAEHAE}

In Section \ref{sec:result_IM_galpop}, we show that the LAEs have the strongest CCF signal among all the subsamples at $r<$ a few $h^{-1}$ Mpc. In Section \ref{sec:result_mukae}, we also tentatively find that the LAEs have a slightly wider $\Sigma_\text{gal}$ distribution than the $\Mstar$--9 and $\Mstar$--8 subsamples, which are comparable in stellar mass to LAEs (e.g., \citealp{hagen14,hagen16,shimakawa17a,kusakabe18,Khostovan19}, see also Figure \ref{fig:hist_Ms}). 
These results suggest that on both small and large-scales, LAEs tend to be located in higher IGM density regions than galaxies with comparably low stellar masses. 
However, our LAEs seem to favor lower stellar masses on average than the other line emitters as shown in Figure~\ref{fig:hist_Ms}\footnote{
We derive $M_\star$ values for LAEs with Spitzer/IRAC photometry by SED fitting in a similar manner to that used in \citet{Kusakabe15,kusakabe18}. The stellar masses of AGNs, HAEs, and O3Es are taken from L16 and S16. 
Note that not all galaxies have  $M_\star$ measurements. Thus, we only plot galaxies with $M_\star$ measurements in Figure~\ref{fig:hist_Ms}.}.
The result that LAEs tend to be found in high IGM density regions despite their low masses apparently conflicts with the hierarchical structure formation model. Although we have not been able to resolve this conflict, we offer two possible explanations below.
One is that regions where LAEs exist are those with low matter
densities but with high {\sc Hi} fractions. Indeed, the IGM {\sc Hi} fraction can vary depending on the presence or absence of strong radiative sources such as starbursts and AGNs (e.g., \citealp{umehata19,mukae19}). It is, however, not clear whether this explanation can also be compatible with the result that the LAEs have the strongest CCF signal among all the galaxy populations.
Another possibility may be that LAEs are satellite galaxies associated with massive halos, although this possibility is apparently incompatible with the observed weak clustering of LAEs at $z=2-3$ (e.g., \citealp{Guaita10,kusakabe18}).
Therefore, this strongest CCF signal of LAEs is puzzling because both the stellar-mass divided subsamples in this paper discussed below and \citet{momose20a} show that lower-mass galaxies correlate more weakly with the IGM.

Another intriguing feature of the LAEs' CCF is its shape, which shows a flat profile until $r\sim3$ $h^{-1}$ Mpc. We find that such a profile cannot be reproduced unless most LAEs are located not in an {\sc Hi} density peak but $2-3$ $h^{-1}$ Mpc away from it. Indeed, such a situation is evident in the postage stamps of LAEs (Figure \ref{fig:shematic} in Appendix~\ref{app:map_delf}). That is to say, LAEs may not faithfully trace the underlying cosmic web. 

When the six subsamples selected on LAE properties are considered (see Figure \ref{fig:IM_LAE}), we find another interesting trend:
that LAEs with faint $L_\text{Ly$\alpha$}$, small $EW_\text{Ly$\alpha$}$, and bright $L_\text{UV}$ have a higher signal than their counterparts with opposite properties.
We argue that its origin is possibly a mass-dependent IGM--galaxy correlation. 
\citet{momose20a} have found that more massive galaxies have a higher CCF signal. 
This result, combined with the fact that LAEs with a smaller $EW_\text{Ly$\alpha$}$ and a brighter $L_\text{UV}$ tend to be more massive (e.g., \citealp{Khostovan19,kusakabe19}) is qualitatively consistent with the trend found in Figure \ref{fig:IM_LAE}.
Note, however, that this explanation is apparently inconsistent with the result that the CCF of LAEs is higher than that of continuum-selected galaxies with similar $\Mstar$. 
On the other hand, LAEs with large $EW_\text{Ly$\alpha$}$ and faint $L_\text{UV}$ show a similar CCF profile to that of continuum-selected galaxies. It may indicate that less massive LAEs trace the IGM distribution in a similar manner to continuum-selected galaxies.

We also discuss a possible contribution of AGNs in our LAE sample based on $L_\text{Ly$\alpha$}$-dependent CCFs. Some observations have suggested that the AGN fraction is close to unity at $L_\text{Ly$\alpha$}\geq10^{43}$ erg s$^{-1}$ (e.g., \citealp{konno16,sobral18a}). Although our LAEs do not have any clear AGN signatures, contamination by hidden AGNs cannot be ruled out. Given that AGNs are hosted by more massive dark matter halos than LAEs, they should show a stronger CCF signal than LAEs, and may have a similar CCF to our AGNs and SMGs (see also Section \ref{sec:dis_AGNSMG}). 
However, because our $L_\text{Ly$\alpha$}$-luminous subsample does not have either a stronger CCF signal or a similar CCF profile to those of AGNs and SMGs in Figure \ref{fig:IM_agns}, the influence of hidden AGNs may be negligible.

Another interesting feature seen in Figure \ref{fig:IM_LAE} is that LAEs with faint $L_\text{Ly$\alpha$}$ and small $EW_\text{Ly$\alpha$}$ have a flat CCF profile. 
If the Ly$\alpha$ emission from LAEs is suppressed by {\sc Hi} in the surrounding IGM \citep{gunn65,haiman02,santos04,Dijkstra07}, LAEs in dense environments must have faint $L_\text{Ly$\alpha$}$ and/or small $EW_\text{Ly$\alpha$}$.
Previous observational studies have suggested a possible reduction of the Ly$\alpha$ escape fraction of galaxies in high-density regions due to high IGM densities (e.g., \citealp{toshikawa16,shimakawa17b,ao17}). 
Therefore, the flat profiles seen in Figure \ref{fig:IM_LAE} may suggest that LAEs in density peaks of the IGM cannot be detected, and thus only LAEs off the peaks where the {\sc Hi} density is not very high are detected.


\subsubsection{HAEs}
\label{sec:dis_HAE}

In Figure \ref{fig:IM_galpop}, we find that the CCF of HAEs is comparable to that of the continuum-selected galaxies. It indicates that these two populations trace the IGM in a similar manner.
The consistency of their CCFs is naturally explained by the fact that the normalized number histogram of $\Mstar$ for our HAEs has a peak at $\Mstar\sim10^{10}$ M$_\odot$ (see Figure~\ref{fig:hist_Ms}), which is in the expected mass range of the continuum-selected galaxies.


\subsubsection{O3Es}

Because our O3Es have a similar $\Mstar$ distribution to those of L16/S16 as found in Figure~\ref{fig:hist_Ms}, they are expected to have a similar CCF to those of the continuum-selected galaxies and HAEs. 
Nonetheless, they have a stronger signal than the continuum-selected galaxies and HAEs as shown in Figure \ref{fig:IM_galpop}, suggesting that they reside in higher-density regions.
Thus, our O3Es might be biased toward higher halo masses.

Further cross-correlation analyses to examine $L_\text{[{\sc Oiii}]$\lambda\lambda5007$}$ and $EW_\text{[{\sc Oiii}]$\lambda\lambda5007$}$ dependence in Figure \ref{fig:IM_O3E} show no significant trend between the CCF signal and $L_\text{[{\sc Oiii}]$\lambda\lambda5007$}$ or $EW_\text{[{\sc Oiii}]$\lambda\lambda5007$}$. 
This implies that the IGM--O3Es connection is generally independent of their properties.
However, because there exists a positive correlation between $L_\text{[{\sc Oiii}]}$ of O3Es and hosting dark halo mass \citep{Khostovan18}, the highest CCF signal in the $EW_\text{[{\sc Oiii}]$\lambda\lambda5007$}\geq300$ {\AA} subsample perhaps indicates that only massive O3Es strongly connect to  high-density {\sc Hi}.


\subsubsection{AGNs and SMGs}
\label{sec:dis_AGNSMG}

The CCFs of AGNs and SMGs have very different shapes from that of star-forming galaxies. Although it is unclear in L16-AGNs (Figure \ref{fig:IM_agns} left), the CCF takes the minimum value not at the center but at $r=5-6$ $h^{-1}$ Mpc in both AGNs and SMGs, indicating that they are typically distributed $5-6$ $h^{-1}$ Mpc away from {\sc Hi} density peaks. Indeed, we confirm that they are mainly found at the outskirts of the cosmic web in Figure \ref{fig:shematic}.

If AGNs and SMGs represent massive galaxies with $M_\text{DH}=10^{11}-10^{13}$ M$_\odot$ (e.g., \citealp{myers07,weib09,Allevato11,Allevato12,allevato14,Allevato19,Hickox12,Koutoulidis13,Plionis18,suh19}), they should be found in high IGM density regions on average. However, the average {\sc Hi} density around them is not so high and is in some cases even lower than the cosmic mean.
It implies {\sc Hi} depletion in several comoving Mpc around them. Because a half of our SMGs are also confirmed as AGNs, such {\sc Hi} depletion is likely caused by the IGM {\sc Hi} photoionization, which is called the proximity effect.
\citet{mukae19} have also suggested that the off-center peak of their mean $\delta_\text{F}$ measurements around QSOs is due to the proximity effect.
QSOs at $z=2-3$ have proximity zones of $r=2-10$ $h^{-1}$ Mpc (e.g., \citealp{dodorico08,uchiyama19}), which is consistent with the peak radii of AGNs' and SMGs' CCFs, thus supporting our interpretation.

The shape and the peak positions of the CCFs also likely depend on AGN type as we already present in Section \ref{sec:result_IM_galpop}. 
We find that IR-identified ($X$-ray-identified) AGNs show positive (negative) $\xi_\text{$\delta$F}$ values at the center. This suggests that IR-identified AGNs are in {\sc Hi} underdense regions, but $X$-ray-identified AGNs are still in overdense regions. Such different environments depending on AGN type are possibly determined by the balance between the baryon accretion rate (mainly gas) to the host galaxy and the IGM {\sc Hi} photoionization rate. 
If the former is higher (lower) than the latter, the {\sc Hi} around the galaxy can become overdense (underdense). 

Obscured AGNs, including SMGs and IR-identified AGNs, are generally hosted by starburst-like and/or young galaxies (e.g., \citealp{Hatziminaoglou10,ichikawa12}). On the contrary, $X$-ray-identified AGNs, often denoted as Type 1 or unobscured AGNs, are suggested to be hosted by more massive halos of $M_\text{DH}=10^{12}-10^{13}$ M$_\odot$ than obscured AGNs \citep{allevato14,suh19}. 
Because the accretion rate is proportional to the halo mass \citep{dekel13}, the accretion rate of $X$-ray-identified AGNs is perhaps higher than the photoionization rate, and thus their surrounding IGM becomes overdense.  
On the other hand, some studies have shown that Type 2 or obscured AGNs might have a higher Eddington ratio than Type 1 AGNs at the same bolometric luminosity, implying relatively higher photoionization rates (e.g., \citealp{lusso12}). 
If this is the case for our IR-identified AGNs and SMGs, and their accretion rates are not high enough to exceed the photoionization rate, they would ionize the surrounding {\sc Hi} and make {\sc Hi} underdense environments.


\subsection{Comparison of the IGM--Galaxy Connection among Galaxy Populations}
\label{sec:dis_comppower}

Figure \ref{fig:IM_all_galpop} shows that the CCF varies among the galaxy populations.
In addition, the population with the strongest CCF signal is different depending on scale (i.e., large or small). 
For large scales over $r\geq5$ $h^{-1}$ Mpc, AGNs and SMGs have the highest CCF signal among all the populations. 
\citet{momose20a} have shown that the higher a galaxy stellar/halo mass is, the stronger a CCF signal is.
In fact, our AGNs are clearly in the higher-$M_\star$ regime than the other galaxy populations (Figure~\ref{fig:hist_Ms}).
In addition, both AGNs and SMGs are known to be hosted by massive halos ($M_\text{DH}=10^{11}-10^{13}$ M$_\odot$: e.g., \citealp{weib09,Allevato11,Allevato12,allevato14,Allevato19,Hickox12,Koutoulidis13,Plionis18,suh19}).
Hence, their highest CCF signal is reasonable. 

Interestingly, AGNs and SMGs are not in high-density regions at small scales within $r=4-5$~$h^{-1}$\,Mpc. 
We argue that it is due to their proximity effect (see also Section \ref{sec:dis_AGNSMG}).
Instead, LAEs show the highest CCF signal among all the populations at small scales, suggesting that they are in the densest {\sc Hi} regions. This result is apparently inconsistent with the fact that LAEs are typically hosted by low-mass halos ($M_\text{DH}=10^{10}-10^{11}$~M$_\odot$, e.g., \citealp{guita11,kusakabe18,Khostovan19}).
This inconsistency is also confirmed in the number histogram of $M_\star$ in Figure~\ref{fig:hist_Ms}, showing that LAEs are biased toward lower $M_\star$ than the other line emitters. In order to identify the reason why LAEs are in higher IGM density regions than the other galaxy populations on small scales, more investigations based on larger galaxy samples are essential.

Another feature of the CCF worth comparing among all the galaxy populations is its shape. 
If a given galaxy population faithfully traces the underlying {\sc Hi} density structure, its CCF should increase toward the cosmic mean ($\xi_\text{$\delta$F}=0$) monotonically. 
All star-forming galaxies except LAEs show such CCFs (Figure \ref{fig:IM_all_galpop}). 
However, LAEs have a flat CCF shape up to $r\sim3$\,$h^{-1}$\,Mpc, suggesting that they are in a few $h^{-1}$\,Mpc away from peaks of the cosmic web. 
This means that overdense regions traced by LAEs do not agree with those traced by other star-forming galaxies. Such a discordance has also been reported in the literature (e.g., \citealp{shimakawa17b,shi19}).
It may be due to the attenuation of Ly$\alpha$ emission by abundant {\sc Hi} at the peaks of cosmic web.


\subsection{Possible Relation between Galaxies and IGM in Terms of Galaxy Evolution}
\label{sec:dis_scenario}

Finally, we discuss how galaxies correlate with the IGM in terms of their evolution by combining all of our results and discussion. 
After their birth, galaxies acquire gas from intergalactic space and stay in the main-sequence while they form stars. 
During this period, the CCF on both large and small scales is determined by the host halo mass of galaxies, as is also indicated in \citet{momose20a}.
According to the theoretical framework of galaxy evolution (e.g., \citealp{hopkins08}), massive galaxies experience the AGN/QSO phase. 
When the AGN activity becomes prominent, galaxies radiate strong ionizing photons and generate a Mpc-scale proximity region, thus suppressing the CCF on small scales as seen for our AGNs and SMGs.
However, because AGNs and SMGs are generally hosted by more massive halos than star-forming galaxies, the total gas density around them on large scales will be higher, as confirmed for our AGNs and SMGs. 
After the AGN and/or QSO phase, galaxies become gradually senescent and quiescent owing to the quenched star formation (e.g., \citealp{hopkins08}). 
The Mpc-scale IGM {\sc Hi} environments may be determined by the balance between accretion rate and {\sc Hi} photoionization rate in the IGM as we discussed in Section~\ref{sec:dis_AGNSMG}. 
Because such galaxies are generally hosted by more massive halos, the large-scale {\sc Hi} density would be possibly high, and even higher than those of AGNs and SMGs. However, we cannot verify the hypothesis from current observational data owing to the lack of large quiescent galaxy samples. 
We leave further investigations for our future work.

\section{summary}

In this study, we investigate the IGM--galaxy connection, paying attention to its dependence on galactic properties, such as $\Mstar$, SFR, sSFR, and their populations.
Using the publicly available 3D Ly$\alpha$ absorption tomography data CLAMATO \citep{lee16,lee18}, and several galaxy catalogs in the literature, we measure the CCF between IGM {\sc Hi} and galaxies and examine the correlation between $\langle \delta_\text{F} \rangle$ and galaxy number density $\Sigma_\text{gal}$. The results of this study are summarized below.

\begin{enumerate}
    \item We detect a CCF signal up to $r\sim50$\,$h^{-1}$\,Mpc from the continuum-selected galaxies (Figure~\ref{fig:IM_specall}). 
    We compare it with those of $\Mstar$, SFR, and sSFR subsamples of simulated galaxies in \citet{momose20a}, and find that the results of $\Mstar$--9, SFR--(iii), SFR--(iv), and sSFR--(ii) subsamples agree with the observed one over $r=3-20$\,$h^{-1}$\,Mpc (Figure~\ref{fig:xi_sim}). 
    In contrast, within the observed galaxies, the CCF of the continuum-selected galaxies agrees with the $\Mstar$--9, $\Mstar$--10, SFR--(ii), SFR--(iii), and sSFR--(i) subsamples. These small discrepancies between the observed and simulated galaxies may be attributed to differences in SED models used in the photo-$z$ catalogs (i.e., L16 and S16) and 
    \citet{momose20a}.
    \item We divide the continuum-selected galaxies into two to four subsamples based on $\Mstar$, SFR, sSFR, and galaxy type (either SFG or QG) measurements given in L16 and S16 and calculate cross-correlations (Figure~\ref{fig:IM_obscomp1}).
    Between L16 and S16, we confirm the consistency of CCFs only in the $\Mstar$--9, $\Mstar$--10, SFR--(ii), sSFR--(i), and SFG subsamples. In addition, we do not confirm the $\Mstar$, SFR, and sSFR dependence on the CCF that is found by \citet{momose20a} for simulated galaxies.
    We suggest that the lack of CCF trends could be a result of a combination of 1) small sample sizes and 2) random and systematic errors in $\Mstar$ and SFR estimates. 
    \item We calculate CCFs for LAEs, HAEs, O3Es, AGNs, and SMGs and obtain the following results. 
    \begin{description}
        \item[$\bullet$~LAEs] LAEs are found to have the strongest CCF signal at the center, and hence reside in the highest-density regions, among all the galaxy populations examined in this study (Figure~\ref{fig:IM_all_galpop}).
        We also find that LAEs with faint $L_\text{Ly$\alpha$}$, small $EW_\text{Ly$\alpha$}$, and bright $L_\text{UV}$ have a stronger CCF signal (Figure~\ref{fig:IM_LAE}).
        We also find the CCF is flat up to $r=3$\,$h^{-1}$\,Mpc. 
        It probably reflects the fact that LAEs do not reside in the density peaks of the IGM, but a few Mpc away from them.         
        Such offsets may be due to the attenuation of Ly$\alpha$ emission by abundant {\sc Hi} in high-density regions of the cosmic web.
        \item[$\bullet$~HAEs] The CCF of HAEs is comparable to that of continuum-selected galaxies (Figure~\ref{fig:IM_galpop}).
        It indicates that these two populations trace the IGM in a similar manner because of similar stellar masses.
        \item[$\bullet$~O3Es] Although we expect similar CCF strengths between HAEs and O3Es considering their comparable stellar masses, the latter have a higher CCF (Figure~\ref{fig:IM_galpop}).
        Because our O3Es with spec-$z$ measurements are biased toward higher [{\sc Oiii}] $\lambda\lambda5007$ luminosities, they may be biased toward higher stellar (and hosting halo) masses.
        \item[$\bullet$~AGNs~$\&$~SMGs]
        AGNs and SMGs commonly have a negative peak at $r\sim5$ $h^{-1}$ Mpc (Figure~\ref{fig:IM_agns}), implying that they tend to be in locally low-density regions. 
        Considering that a half of our SMGs are also confirmed as AGNs, such {\sc Hi} depletion may be due to the proximity effect.
        We also find a hint that the CCF of IR ($X$-ray) identified AGNs is weaker (stronger) at the center. This difference may imply that IR identified AGNs have higher photoionization rates.
    \end{description}
    \item On large scales ($r\geq5$ $h^{-1}$ Mpc), AGNs and SMGs have the highest CCF amplitude among all the populations. This is reasonable because they are generally hosted by the most massive halos with $M_\text{DH}=10^{11}-10^{13}$ M$_\odot$.
    On small scales ($r<5$\,$h^{-1}$\,Mpc), on the other hand, LAEs show the highest signal.
    However, the cause of such a high signal in LAEs, which are typically hosted by low-mass halos, is still unclear (see Figure~\ref{fig:IM_all_galpop}). 
    \item We examine the correlation between $\langle \delta_\text{F} \rangle$ and $\Sigma_\text{gal}$ (``overdensity analysis''; see Figures~\ref{fig:den_bias} and \ref{fig:den_unbias}). We only confirm statistically significant anti-correlations in the L16-$\Mstar$--9 and ALL subsamples. Their slopes are comparable to that in the literature but steeper than those found in \citet{momose20a}, probably due to photo-$z$ errors.
    We also tentatively find that LAEs have a slightly wider $\Sigma_\text{gal}$ distribution than the L16-$\Mstar$--9 and L16-$\Mstar$--8 subsamples at the same redshift slice, which are comparable in stellar mass to LAEs. It may suggest that LAEs have a stronger correlation with the IGM {\sc Hi} for their stellar masses.
\end{enumerate}

\section*{Acknowledgements}

We appreciate the anonymous referee for careful reviewing and useful comments that improved our paper. 
We are grateful to Dr. K.-G. Lee for providing the CLAMATO data and Dr. D. Sobral for providing the mask data of \citet{sobral13}. 
We thank Drs. M. Rauch, F. S. Zahedy, K. Ichikawa, T. Kawamuro, M. Imanishi, H. Yajima, D. Sorini, T. Suarez Noguez, K. Kakiichi, and R. A. Meyer for helpful discussions. 
R.M. acknowledges a Japan Society for the Promotion of Science (JSPS) Fellowship at Japan. 
This work is supported by the JSPS KAKENHI grant Nos. JP18J40088 (RM), JP19K03924 (KS), and JP17H01111, 19H05810 (KN). 
The \citet{laigle16} galaxy catalog is based on data products from observations made with ESO Telescopes at the La Silla Paranal Observatory under ESO program ID 179.A-2005 and on data products produced by TERAPIX and the Cambridge Astronomy Survey Unit on behalf of the UltraVISTA consortium. We acknowledge the Python programming language and its packages of numpy, matplotlib, scipy, and astropy \citep{astropy13}.


\appendix

\section{$\langle\Delta_\text{F}\rangle$ maps}
\label{app:map_delf}
\begin{figure*}
	\begin{center}
	\includegraphics[width=0.95\linewidth]{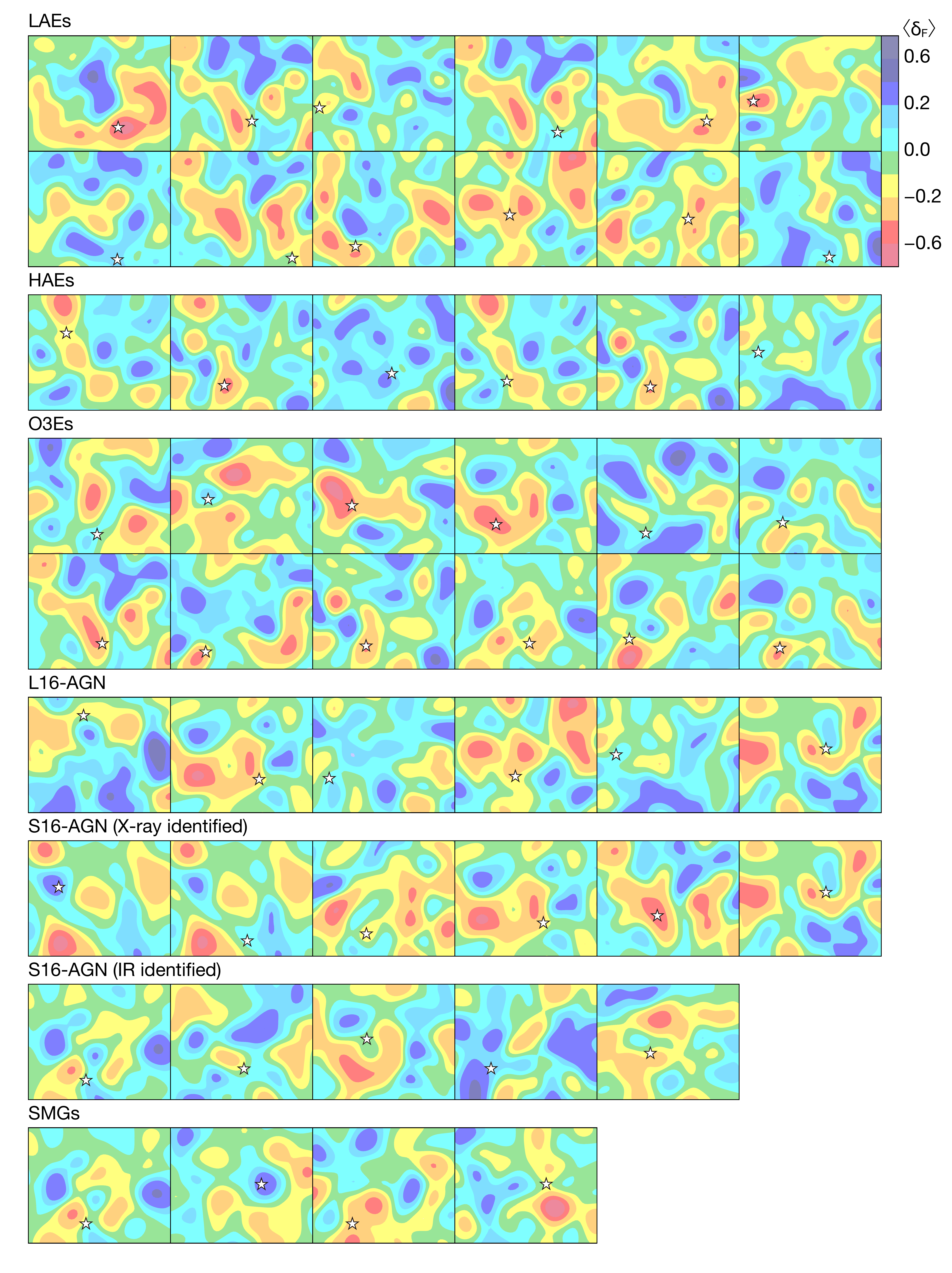} 
	\caption{
    Projected {\sc Hi} density maps ($30$~$h^{-1}$~Mpc~$\times$~$24$~$h^{-1}$~Mpc) for randomly selected galaxies from each subsample, obtained by collapsing thin ($\Delta z=2$ $h^{-1}$ Mpc) data cubes around them. Open stars indicate the position of galaxies. Warm and cold colors denote overdense and underdense regions, respectively.
    Density maps of LAEs, HAEs and O3Es are arranged in the order of line luminosity from left to right, spanning two rows in the cases of LAEs and O3Es. 
    }
	\label{fig:shematic}
	\end{center}
\end{figure*}

In order to visualize the IGM {\sc Hi} density fluctuations around galaxies used in the CCF analysis, we make postage stamp images of the projected {\sc Hi} density distribution by collapsing a thin ($\Delta z=2$ $h^{-1}$ Mpc) CLAMATO cube centered at each galaxy. Selected examples of each galaxy population are shown in Figure \ref{fig:shematic}. The galaxy position on the sky is marked by a white star. 


\section{Mass, SFR, sSFR dependence on the CCFs}
\label{app:IM_div2}

\begin{figure*}
	\begin{center}
	\includegraphics[width=\linewidth]{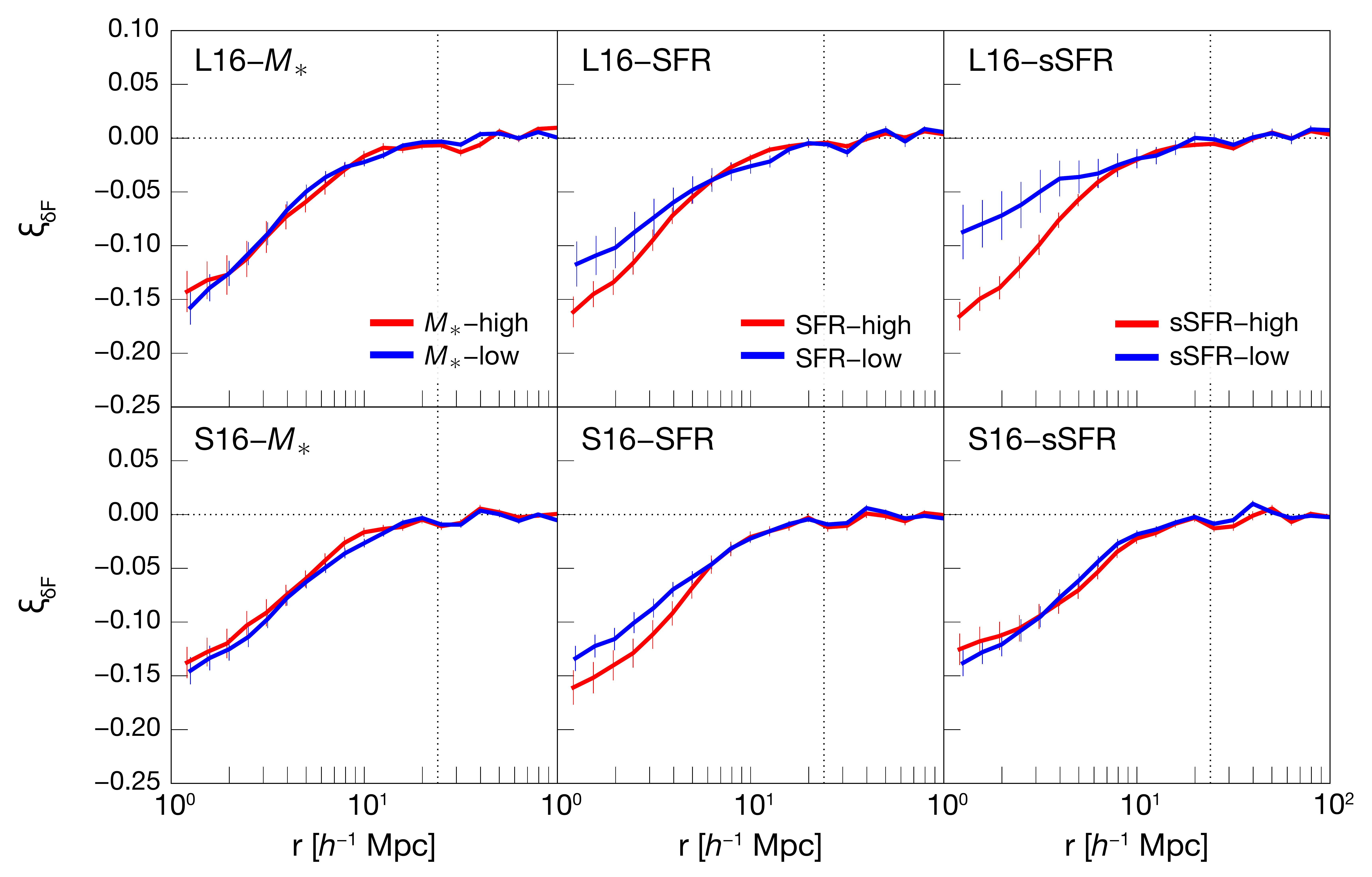} 
	\caption{
	CCFs of (top) L16 and (bottom) S16 subsample pairs divided by $\Mstar$, SFR, and sSFR from left to right as a function of radius in comoving units.
    The meaning of vertical dotted lines is the same as in Figure \ref{fig:IM_specall}.}
	\label{fig:IM_obscomp2}
	\end{center}
\end{figure*}

We find no significant dependence of the CCF on galactic properties in Section~\ref{sec:result_IM_galpro} and Figure~\ref{fig:IM_obscomp1}, in contrast to what \citet{momose20a} have found for simulated galaxies.
This lack of dependence could have resulted from large statistical uncertainties owing to the small sample sizes.
To reduce the statistical uncertainties, we also perform a similar analysis by splitting L16 and S16 into only two subsamples by $\Mstar$, SFR, and sSFR (Figure~\ref{fig:IM_obscomp2}). Note that we use $\Mstar=10^{10}$ M$_\odot$, SFR $=10^{1}$ M$_\odot$ yr$^{-1}$, and sSFR $=10^{-9}$ yr$^{-1}$ as the border. 
We find that, for both L16 and S16, the higher-SFR subsample has a higher CCF at the $2\sigma$ significance level up to $r\sim6$~$h^{-1}$~Mpc. Although it is in qualitative agreement with the trend found in \citet{momose20a}, the statistical significance may not be high enough to confirm the trend.
On the other hand, no significant (or consistent) dependence is found on either $\Mstar$ or sSFR.
A more precise analysis requires much larger galaxy samples in each category.


\bibliographystyle{aasjournal}
\bibliography{clam}

\end{document}